 \documentclass[authoryear,preprint,12pt]{elsarticle}

\usepackage{amssymb}
 \usepackage{amsmath}
  \usepackage{algorithm}
 \usepackage{algpseudocode}
 \usepackage{booktabs}

\usepackage{natbib,geometry,fleqn,graphicx,hyperref}
\usepackage{subfigure,color,colordvi,xcolor}

\journal{Computer-Aided Design}

\newtheorem{remark}{Remark}

\newcommand{\bm}[1]{\text{\boldmath $#1$\unboldmath}}

\newcommand{\nsd}  {\ensuremath{\texttt{n}_{\texttt{sd}}}}
\newcommand{\bx}{\bm{x}}

\newcommand{\bv}{\bm{v}}

\newcommand{\nel}{\ensuremath{\texttt{n}_{\texttt{el}}}}
\newcommand{\nelB}{\nel}
\newcommand{\nno}{\ensuremath{\texttt{n}_{\texttt{no}}}}
\newcommand{\nnoB}{\nno}

\newcommand{\metric}{\bm{\mathcal{M}}}
\newcommand{\be}{\mathbf{e}}
\newcommand{\bR}{\mathbf{R}}
\newcommand{\bP}{\mathbf{P}}

\newcommand{\param}{\vartheta}

\DeclareMathOperator{\diag}{diag}

\begin{document}

\begin{frontmatter}
\title{Anisotropic mesh spacing prediction using neural networks}

\author{Callum Lock}
\ead{C.D.Lock@Swansea.ac.uk}

\author{Oubay Hassan}
\ead{O.Hassan@Swansea.ac.uk}
\author{Ruben Sevilla\corref{cor1}}
\ead{R.Sevilla@Swansea.ac.uk}
\author{Jason Jones}
\ead{J.W.Jones@Swansea.ac.uk}

\cortext[cor1]{Corresponding author}

\affiliation{
		organization={Zienkiewicz Institute for Modelling, Data and AI, Faculty of Science and Engineering, Swansea Univeristy},
		city={Swansea},
		postcode={SA1 8EN},
		state={Wales},
		country={UK}
	}

\begin{abstract}
This work presents a framework to predict near-optimal anisotropic spacing functions suitable to perform simulations with unseen operating conditions or geometric configurations. The strategy consists of utilising the vast amount of high fidelity data available in industry to compute a target anisotropic spacing and train an artificial neural network to predict the spacing for unseen scenarios. The trained neural network outputs the metric tensor at the nodes of a coarse background mesh that is then used to generate meshes for unseen cases. Examples are used to demonstrate the effect of the network hyperparameters and the training dataset on the accuracy of the predictions. The potential is demonstrated for examples involving up to 11 geometric parameters on CFD simulations involving a full aircraft configuration.
\end{abstract}


\end{frontmatter}

%%%%%%%%%%%%%%%%%%%%%%%%%%%%%%%%%%%%%%%%%%%%%%%%%%%%%%%%%%%%%%%%%%%%%%%%%%%%%%%%%%%%%%%%%%%%%%%%%%%%%%%
\section{Introduction}\label{sc:intro}
%%%%%%%%%%%%%%%%%%%%%%%%%%%%%%%%%%%%%%%%%%%%%%%%%%%%%%%%%%%%%%%%%%%%%%%%%%%%%%%%%%%%%%%%%%%%%%%%%%%%%%%

Unstructured mesh generation plays a critical role in numerical simulations, influencing both computational efficiency and solution accuracy. Traditional unstructured mesh generation techniques often rely on heuristic-based refinement criteria and expert-driven adjustments, which can be computationally expensive and time-consuming~\cite{dawes2001reducing,slotnick2014cfd,karman2017mesh}. In fact, the generation of unstructured meshes for complex geometries continues to be one of the most time-consuming stages in the simulation process. The problem is exacerbated when multiple simulations for different operating conditions or geometric configurations are required, as usual in a design or optimisation cycle.

One obvious alternative, when the same geometry is to be analysed for different operating conditions, is to generate a single mesh, fine enough to capture all the features for all the solutions to be computed~\cite{MichalTetrahedron}. This practice was analysed in~\cite{lock2023meshing} in terms of computational time and carbon emissions associated with the use of HPC facilities and it was concluded that it can lead to 35 times more carbon emissions when compared to running simulations on meshes tailored for each simulation.

Another alternative is to use automatic mesh adaptivity~\cite{meshECM}. These methods require only an initial mesh, which is iteratively refined based on error estimation. However, the effectiveness of the adaptation depends on the initial mesh capturing key solution features. Moreover, single-entity error indicators often miss complex phenomena, necessitating multiple indicators that incorporate both primitive and derived variables to detect features like shock waves, contact discontinuities, flow separation, and turbulence-related structures. In addition, for complex three-dimensional simulations, the full adaptation process can involve 20 to 30 iterations~\cite{loseille2010fully}, with each cycle requiring solution computation, error estimation, mesh refinement, and solution interpolation.

Recent advancements in machine learning, particularly neural networks, have demonstrated their potential to enhance mesh generation by automating key aspects of the process. The use of machine learning to assist the mesh generation process was originally proposed in~\cite{chang1991self} but it was not until the early 2000s that the first examples involving three dimensional examples emerged~\cite{alfonzetti2003neural}, still involving simple geometries and in the context of electromagnetic problems. More recently the use of machine learning techniques for mesh generation and adaptation has received increasing attention. 

In~\cite{zhang2021meshingnet3d}, artificial neural networks (ANNs) were used to predict spacing at a given location based on parameters related to partial differential equations, geometry, and boundary conditions. A similar approach for computational fluid dynamics (CFD) was proposed in~\cite{huang2021machine}, where spacing from adaptively refined meshes is mapped onto a Cartesian grid, converted into a greyscale image, and used to train an ANN. These methods relate to the present work as they predict spacing functions for mesh generation in unseen cases. Other studies have explored ANNs for mesh and degree adaptivity~\cite{yang2021reinforcement,wallwork2022e2n,tlales2022machine}, predicting near-optimal spacing for complex CFD applications~\cite{lock2023meshing,lock2023predicting}, and estimating mesh anisotropy~\cite{fidkowski2021metric}.

In this work a new approach is presented to predict an anisotropic spacing function for new simulations involving either new operating conditions or new geometric configurations. The strategy follows the rationale introduced in~\cite{lock2023predicting} as it uses a background mesh to define a discrete spacing field. However, the current work extends the previous work to enable the prediction of a metric tensor rather than predicting a scalar. 

First, a strategy is proposed to compute the metric tensor at each point of a computational mesh that was used to perform a high-fidelity simulation. Second, a method to transfer the anisotropic spacing to a coarse background mesh is presented. The strategy introduces mesh morphing based on a classical Delaunay graph to ensure that the spacing can be transferred to the same background mesh, even when the simulations are performed for different geometries. Finally, a new ANN is proposed to enable the prediction of the metric tensor at each point of the background mesh for new, unseen, flow conditions or geometric configurations. The proposed ANN architecture exploits the mathematical properties of the metric tensors to ensure that the prediction leads to a valid metric tensor and also to minimise the required amount of outputs to be predicted. 

Numerical results involve three dimensional examples in the context of inviscid compressible flows with variable flow conditions and geometric parameters. The examples, involving up to 11 geometric parameters demonstrate the potential and the accuracy of the proposed approach to define a near-optimal initial mesh that is suitable to perform simulations. It is worth noting that the present approach can be seen as a way to obtain a good initial mesh for new simulations and, if necessary, it can be integrated in a mesh adaptive cycle. If used in this context, the predicted mesh is expected to require very few, if any, adaptivity cycles to reach the required accuracy. In fact, numerical examples included in this work assess the suitability of the meshes to perform simulations.

The remainder of the paper is organised as follows. Section~\ref{sc:background} presents the necessary concepts to introduce the proposed approach. These concepts are the definition of an anisotropic spacing field using a background mesh, the ANNs and the mesh morphing approach for variable geometries. Section~\ref{sc:near-optimal} contains the main scientific contribution of this work, including the strategy to identify the anisotropic spacing that is required to generate a mesh suitable to capture a given solution, the strategy to transfer this information to a coarse background mesh and the proposed ANN model. Particular attention is paid to the mathematical properties of the metric tensor to be predicted. In Section~\ref{sc:examples} two numerical examples are presented. The first example involves the geometry of the ONERA M6 wing with variable free-stream Mach number and angle of attack, whereas the second example considers a full aircraft configuration parametrised with 11 geometric parameters. A study of the suitability of the predicted meshes to perform accurate simulations is also included. Finally, Section~\ref{sc:conclusions} summarises the conclusions of the work that has been presented.

%%%%%%%%%%%%%%%%%%%%%%%%%%%%%%%%%%%%%%%%%%%%%%%hh JJ%%%%%%%%%%%%%%%%%%%%%%%%%%%%%%%%%%%%%%%%%%%%%%%%%%%%%%%%
\section{Background}\label{sc:background}
%%%%%%%%%%%%%%%%%%%%%%%%%%%%%%%%%%%%%%%%%%%%%%%hh JJ%%%%%%%%%%%%%%%%%%%%%%%%%%%%%%%%%%%%%%%%%%%%%%%%%%%%%%%%

This section outlines the fundamental concepts necessary to introduce the proposed approach for predicting near-optimal anisotropic spacing, suitable for generating meshes of previously unseen simulations.

%-------------------------------------------------------------------------------
\subsection{Anisotropic mesh spacing and control using a background mesh}\label{sc:meshing}
%-------------------------------------------------------------------------------

The generation of suitable meshes for simulation often requires a significant level of human expertise and interaction to define a suitable spacing function. Among all the available techniques to define a spacing function, the use of mesh sources and a background mesh provides significant flexibility as it enables the definition of a spacing function that is independent of geometric entities. 

This work considers the use of a background mesh for three main reasons. First, previous work concerning the use of ANNs for predicting near-optimal isotropic mesh spacing~\cite{lock2023meshing,lock2023predicting,sanchez2024machine} has shown that, compared to the prediction of mesh sources, predicting the spacing at a background mesh leads to higher accuracy for the same amount of available training data. Second, predicting the spacing at a background mesh leads to faster training times given the reduced number of outputs of the ANN, compared to the prediction of a set of mesh sources. Finally, using a background mesh it is easier to define anisotropic spacing functions, which is the main focus on this work, contrary to previous work which was restricted to isotropic spacing.

The process to specify a spacing function starts by creating a background mesh, $\mathcal{B}_h$, defined as a collection of $\nelB$ disjoint elements $\{B_e\}_{e=1,\ldots,\nelB}$ that covers the computational domain $\Omega$, i.e. $\Omega \subseteq \cup_{e=1}^{\nelB} B_e$. In this work the background mesh is generated automatically using curvature control to provide local refinement near the regions of interest.

When isotropic spacing is of interest, the spacing function is simply defined as a nodal field in the background mesh, namely $\{\delta_i\}_{i=1,\ldots,\nnoB}$, where the number of background mesh nodes is denoted by $\nnoB$. Once a spacing function is available, the generation of a computational mesh requires the computation of the desired spacing at a point $\bx \in \Omega$. This operation is performed by identifying the element of the background mesh that contains the point $\bx$ and using the nodal values of the identified element to interpolate the spacing from the nodal values. In this work the background mesh is always an unstructured tetrahedral mesh so the computation of the spacing at a point uses a linear interpolation of the four nodal values of the identified element.

The definition of an anisotropic spacing function in a background mesh is achieved by associating to each point a metric tensor~\cite{thompson1998handbook}, which encapsulates the information of three mutually orthogonal directions, $\be_1$, $\be_2$ and $\be_3$, and the desired spacing in each direction, $\delta_1$, $\delta_2$ and $\delta_3$. The metric tensor at a point is formally defined as
\begin{equation}\label{eq:metric}
\metric = \bR \bm{\Lambda}^{2} \bR^{T},
\end{equation}
where the columns of $\bR$ are the vectors $\be_1$, $\be_2$ and $\be_3$ and $\bm{\Lambda}^{2} =\diag\left(1/\delta_1^2,1/\delta_2^2,1/\delta_3^2\right)$. To simplify the presentation it is assumed that the spacings are ordered, i.e. $\delta_1 \leq \delta_2  \leq \delta_3$.

\begin{figure}[!tb]
	\centering	
	\includegraphics[width=0.6\textwidth]{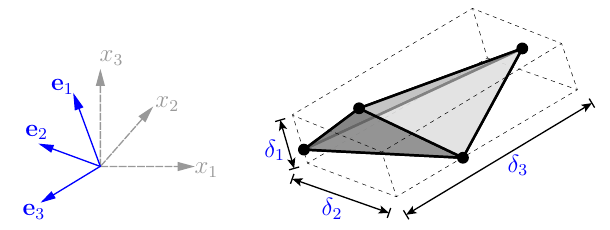}
	\caption{Representation of a tetrahedral element with anisotropic spacing given by three mutually orthogonal directions, $\be_1$, $\be_2$ and $\be_3$, and the desired spacing in each direction, $\delta_1$, $\delta_2$ and $\delta_3$.}
	\label{fig:ElementDefinition}
\end{figure}

The definition of an element with anisotropic spacing is illustrated in Figure~\ref{fig:ElementDefinition}.

\begin{figure}[!tb]
	\centering	
	\includegraphics[width=0.3\textwidth]{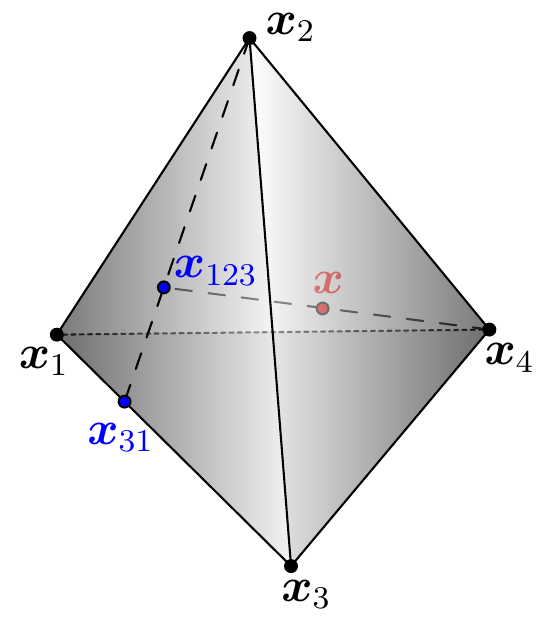}
	\caption{Schematic representation of the metric interpolation for a point inside a tetrahedral element.}
	\label{fig:MetricInterpolationElem}
\end{figure}

The generation of an anisotropic mesh requires the definition of a suitable metric tensor at a point in the computational domain, $\bx \in \Omega$, from the metric tensors defined at the nodes of the element in the background mesh that contains $\bx$. To this end, a metric interpolation strategy is commonly adopted~\cite{frey2007mesh}. In this work, the interpolation of the metric between two points $\bx_1$ and $\bx_2$ with metrics $\metric_1$ and $\metric_2$ respectively, is defined as
\begin{equation}\label{eq:metricInterpolation}
\metric(t) = \left( (1 - t) \, \metric_1^{-\frac{1}{2}} + t \, \metric_2^{-\frac{1}{2}} \right)^{-2},
\quad \text{for $t \in [0,1]$}.
\end{equation}
The interpolation of a metric in a generic point $\bx$ contained in an element of the background mesh with nodes $\bx_1$, $\bx_2$, $\bx_3$ and $\bx_4$ is performed as follows. First, the intersection of the line connecting $\bx_4$ and $\bx$ with the triangular face formed by nodes $\bx_1$, $\bx_2$ and $\bx_3$, denoted as $\bx_{123}$, is computed. Next, the intersection of the line connecting $\bx_2$ and $\bx_{123}$ with the edge connecting nodes  $\bx_3$ and $\bx_1$, denoted as $\bx_{31}$, is computed. The metric interpolation is applied to the metrics at nodes $\bx_1$ and $\bx_3$ to compute the metric at $\bx_{31}$. Then the metric interpolation is applied to the metrics at nodes $\bx_2$ and $\bx_{31}$  to compute the metric at $\bx_{123}$. Finally the metric interpolation is applied to the metrics at nodes $\bx_4$ and $\bx_{123}$  to compute the desired metric at $\bx$. The process is illustrated in Figure~\ref{fig:MetricInterpolationElem}.

%-------------------------------------------------------------------------------
\subsection{Feed-forward neural networks}\label{sc:NNs}
%-------------------------------------------------------------------------------

A feed-forward artificial neural network (ANN) consists of neurons arranged in layers, where each neuron is connected only to all the neurons in the preceding and succeeding layers. Each connection between two neurons is assigned a weight, and each layer is given a bias, which can be considered an additional neuron with a fixed value. The first layer comprises the ANN inputs, while the neurons in the final layer are referred to as outputs. The intermediate layers, known as hidden layers, are typically indexed as $l=1,\ldots...,N_{l}$, where $N_{l}$ denotes the total number of hidden layers~\cite{hagan1997neural}.

\begin{figure}[!tb]
	\centering
	\includegraphics[width=0.8\textwidth]{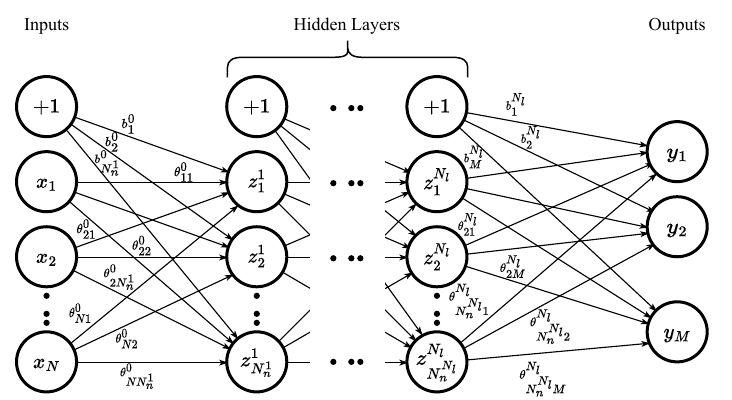}
	\caption{Schematic representation of a feed-forward ANN.}
	\label{fig:NN}
\end{figure}

Forward propagation involves computing the value associated with each neuron based on the values of the neurons in the preceding layer, the weights of the corresponding connections, and an activation function. Specifically, the value of the $j$-th neuron of the $l+1$ layer is computed as
\begin{equation}\label{eq:neuronEq}
	z^{l+1}_{j} = F^{l+1} \left( \sum_{i=1}^{N_n^l}\theta^{l}_{ij}z^{l}_{i} + b^{l}_{j} \right),
\end{equation}
where $F^l$ is the activation function of layer $l$, $\theta^{l}_{ij}$ is the weight o the connection between the $i$-th neuron in layer $l$ and the $j$-th neuron in layer $l+1$, $b^l$ is the bias of layer $l$, and $N_n^l$ is the number of neurons in layer$l$. Figure~\ref{fig:NN} illustrates a schematic representation of a feed-forward ANN.

In this work two types of ANNs are considered. One ANN is used to predict the spacing associated to the three mutually orthogonal directions that define a metric tensor whereas a different ANN is used to predict the three directions. For the first ANN, the error or cost function is defined as
\begin{equation} \label{eq:costFunction}
	E_\delta(\bm{\theta},\mathbf{b}) = \frac{1}{N_{tr}M} \sum_{k=1}^{N_{tr}} \sum_{i=1}^M \left[ y^k_i(\mathbf{x}^k) - h_i^k(\mathbf{x}^k,\bm{\theta},\mathbf{b}) \right]^2,
\end{equation}
where the output predicted by the ANN is denoted by $h_i^k(\mathbf{x}^k,\bm{\theta},\mathbf{b})$, $N_{tr}$ is the number of training cases and the inputs and outputs are arranged in two vectors $\mathbf{x}^k = \{x_{1},...,x_{N} \}^T$ and $\mathbf{y}^k = \{y_{1},...,y_{M} \}^T$, respectively, for $k=1,\ldots,N_{tr}$.

When the ANN is built to predict the three orthogonal directions, a measure of the alignment between the predicted and target vectors is preferred. To this end, this work proposes the following error function
\begin{equation} \label{eq:costFunctionVect}
	E_e(\bm{\theta},\mathbf{b}) = \frac{1}{N_{tr}M} \sum_{k=1}^{N_{tr}} \sum_{i=1}^M  \left[1 -  \textbf{y}^k_i(\mathbf{x}^k) \cdot \textbf{h}_i^k(\mathbf{x}^k,\bm{\theta},\mathbf{b})  \right]^2,
\end{equation}
where the target and predicted outputs are now vectors and the deviation of the dot product with respect to one is taken as a measure of alignment between target and predicted directions.

The process of training the ANN consists of finding the weights and biases that minimise the error function. To this end, the ADAM optimiser~\cite{kingma2014adam} is employed in this work, with the parameters selected as in~\cite{lock2023meshing,lock2023predicting}. 

As is common in the context of ANNs, the choice of hyperparameters, such as the number of neurons and hidden layers, can significantly impact the accuracy of predictions. In the numerical examples considered, a simple grid search is performed to determine the optimal hyperparameters. To complete the description of the ANN architecture, the activations functions employed need to be detailed. In this work, the sigmoid function is applied in all hidden layers, while a linear function is used in the output layer, namely
\begin{equation} \label{eq:activation}
F^l(z) = \begin{cases}
	z & \text{if $l=N_l$,}  \\
		\displaystyle \frac{1}{1+e^{-z}} & \text{otherwise.}  \\
\end{cases}
\end{equation}

%-------------------------------------------------------------------------------
\subsection{Mesh morphing for geometrically parametrised domains}\label{sc:morphing}
%-------------------------------------------------------------------------------

When considering geometrically parametrised domains, the training cases correspond to available simulations with different geometries. To transfer the spacing function to the same background mesh, as required to train a simple feed-forward ANN, this work proposes the use of a simple mesh morphing approach based on the Delaunay graph.

Given a set of $N_g$ geometric parameters $\bm{\param} = \{\param_1, \param_2, \ldots, \param_{N_g} \}$ with a pre-defined range for each parameter, $[\param_i^a, \param_i^b]$ for $i=1,\ldots,N_g$, the background mesh is generated for the geometric configuration with parameters $\bar{\bm{\param}} = \{\bar{\param}_1, \bar{\param}_2, \ldots, \bar{\param}_{N_g} \}$, where $\bar{\param}_i = (\param_i^a+\param_i^b)/2$.

A Delaunay tetrahedralisation is built using only the nodes of the background mesh on the boundary of the computational domain. For each mesh node the element of the Delaunay graph that contains that node and its area coordinates are computed and stored. 

For a new geometric configuration the boundary mesh nodes of the background mesh are projected to the new geometry and the new position of each mesh node is computed by evaluating the position corresponding to the pre-stored area coordinates in the original configuration.

The proposed strategy enables the morphing of the same background mesh to match the geometry of all the training cases available, ensuring that all cases share the same number of outputs as required to train a simple feed-forward ANN.

%%%%%%%%%%%%%%%%%%%%%%%%%%%%%%%%%%%%%%%%%%%%%%%%%%%%%%%%%%%%%%%%%%%%%%%%%%%%%%%%%%%%%%%%%%%%%%%%%%%%%%%
\section{Near-optimal anisotropic mesh spacing prediction}\label{sc:near-optimal}
%%%%%%%%%%%%%%%%%%%%%%%%%%%%%%%%%%%%%%%%%%%%%%%%%%%%%%%%%%%%%%%%%%%%%%%%%%%%%%%%%%%%%%%%%%%%%%%%%%%%%%%

This section presents a new approach to predict the anisotropic spacing function on a background mesh that can be used to generate meshes suitable for unseen simulations that involve either new geometric configurations or new flow conditions. It is assumed that the data for training is available from historical high fidelity analysis. 

The proposed methodology can be summarised in the following stages:
\begin{enumerate}
	\item For each solution that is available as training data, identify the anisotropic spacing that could be used to generate a mesh capable of accurately capturing the given solution. 
	\item For each available training case, transfer the anisotropic spacing to a coarse background mesh. This step involves the morphing of the background mesh when geometric parameters are considered.
	\item Train an ANN to predict the metric tensor that defines the anisotropic spacing at each node of the background mesh. 
\end{enumerate}

%============================================================================================
\subsection{Computation of the target anisotropic spacing in the computational mesh} \label{sc:targetSpacingComp}
%============================================================================================

Building on established principles in error analysis~\cite{peraire1992adaptive}, the determination of a discrete spacing function from a given solution involves computing the Hessian matrix of a selected key variable at each point within the computational mesh. This work considers the generation of meshes for inviscid compressible flow simulations so the selected key variable is the pressure, $p$.

The spacing at a node $\bx_i$ in a given direction, defined by a unit vector $\bm{\beta}$, depends on the derivatives of the pressure.
\begin{equation} \label{eq:errorAnalysis}
	\delta_{i,\bm{\beta}}^2 \: \left ( \sum_{k,l=1}^{\nsd} (H_i)_{kl} \beta_k \beta_l \right ) = K,
\end{equation}
where $\nsd$ denotes the number of spatial dimensions, $K$ is a user-specified constant linked to the level of refinement introduced around areas with high gradient and 
\begin{equation} \label{eq:hessian}
	(H_i)_{kl} = \frac{ \partial^2 p_i} {\partial x_k \partial x_l}
\end{equation}
denote the components of the Hessian at a node $\bx_i$. The components of the Hessian matrix are evaluated as the derivatives of the derivatives, and the computation of a derivative is performed using a standard recovery postprocess~\cite{zienkiewicz1992superconvergent1,zienkiewicz1992superconvergent2}.

Once these derivatives have been evaluated, the three mutually orthogonal directions at $\bx_i$ are given by the eigenvectors of the Hessian matrix, $\mathbf{H}_i$, and the spacing in each of those directions is determined as
\begin{equation} \label{eq:spacingMin}
	\delta_{i,j}  =
	\begin{cases}
		\delta_{min} & \text{if $\lambda_{i,j} > K/\delta_{min}^2$,} \\
		\delta_{max} & \text{if $\lambda_{i,j} < K/\delta_{max}^2$,} \\
		\sqrt{ K/\lambda_{i,j}} & \text{otherwise},
	\end{cases}
\end{equation}
where $\{\lambda_{i,j}\}_{j=1,\ldots,\nsd}$ are the eigenvalues of $\mathbf{H}_i$.

Equation~\eqref{eq:spacingMin} ensures that the spacings $\delta_{i,j}$ remain within a user-defined range $[\delta_{min},\delta_{max}]$. The lower bound, $\delta_{min}$, prevents excessive local refinement in regions with steep gradients, such as areas near strong shocks. Similarly, the upper bound, $\delta_{max}$, prevents the spacing from becoming too large in regions of undisturbed flow, where the pressure remains smooth or nearly constant.

The value of $K$ appearing in Equation~\eqref{eq:spacingMin} is defined as 
\begin{equation} \label{eq:spacingK}
	K = S^2 \delta_{min}^2 \lambda_{max}
\end{equation}
where the scaling factor $S \in (0,1]$ is defined by the user and taken as 0.2 in all the examples. Further details about the effect of altering the value of $S$ are discussed in~\cite{sanchez2024machine}.

In addition, in this work, the maximum stretching is restricted to be less than or equal to five. This is based on previous experience that shows that a stretching larger than five leads to an increased number of iterations to converge the solution to steady state with the vertex-centred finite volume solver employed in this work.

%============================================================================================
\subsection{Transfer of the target anisotropic spacing to a background mesh} \label{sc:transferSpacing}
%============================================================================================

Given that the different simulations available as training cases might have been performed in different meshes and a feed-forward ANN requires the same number of outputs for all cases, the proposed strategy involves transferring the spacing function of each simulation to the same background mesh. This approach also aims at reducing the number of outputs in the ANN, and consequently the training time, as the available simulations might have been performed in extremely fine meshes.

In~\cite{lock2023predicting,sanchez2024machine} the authors proposed a \textit{conservative interpolation} approach to perform this task when the spacing function is isotropic, which is briefly summarised here. Given a node, $\bx_i^\texttt{B}$, of the background mesh $\mathcal{B}_h$, the patch of elements that contains the node $\bx_i^\texttt{B}$ is denoted by $\mathcal{P}_i^\texttt{B}$. The subset of nodes of the computational mesh, corresponding to a training case, that are within the patch $\mathcal{P}_i^\texttt{B}$ is denoted by $\mathcal{X}_{\mathcal{P}_i}$. With this notation, the conservative interpolation of the spacing involves defining the spacing at $\bx_i^\texttt{B}$ as the minimum of the spacing of all nodes in $\mathcal{X}_{\mathcal{P}_i}$, namely
\begin{equation} \label{eq:conservativeIntep}
	\delta_i^\texttt{B} = \min_{j \in \mathcal{X}_{\mathcal{P}_i}}\{\delta_j \}.
\end{equation}

To transfer the target anisotropic spacing, computed following the strategy described in the previous section, from a computational mesh to a background mesh, this work proposes a \textit{conservative metric intersection}. Metric intersection~\cite{} is commonly used to find a representative metric at a point when two (or more) metrics are available. Given two metrics $\metric_1$ and $\metric_2$, the metric intersection is formally defined as
\begin{equation}\label{eq:metricInter}
\metric_1 \cap \metric_2 = \bP^{-T} \bm{\Lambda}_{\max} \bP^{-1},
\end{equation}
where $\bm{\Lambda}_{\max} =\diag\left(\max\{\lambda_1,\mu_1\},\max\{\lambda_2,\mu_2\},\max\{\lambda_3,\mu_3\}\right)$ with $\lambda_i = \be_i^T \metric_1 \be_i$ and $\mu_i= \be_i^T \metric_2 \be_i$, for $i=1,\ldots,\nsd$, and  the columns of $\bP$ contain the eigenvectors of $\metric_1^{-1} \metric_2$.

The proposed conservative metric intersection to transfer an anisotropic spacing from a computational mesh to a background mesh is described next.  Given a node, $\bx_i^\texttt{B}$, of the background mesh $\mathcal{B}_h$, the patch of elements that contains the node $\bx_i^\texttt{B}$, denoted by $\mathcal{P}_i^\texttt{B}$, is considered. The subset of nodes of the computational mesh that are within the patch $\mathcal{P}_i^\texttt{B}$ is denoted by $\mathcal{X}_{\mathcal{P}_i}$. With this notation, the conservative metric intersection involves defining the metric for node $\bx_i^\texttt{B}$ as 
\begin{equation} \label{eq:metricInterMultiple}
	\metric_i^\texttt{B} = \bigcap_{j \in \mathcal{X}_{\mathcal{P}_i}} \metric_j.
\end{equation}

\begin{remark}
As reported in~\cite{sanchez2024machine}, special attention must be paid to two special cases. First, if the patch of a background mesh node, $\mathcal{P}_i^\texttt{B}$, does not contain any node of the computational mesh, i.e $\mathcal{P}_i^\texttt{B} = \emptyset$, then the proposed strategy is to utilise the metrics of the nodes of the computational mesh that contain the current background mesh node. Second, for domains with curved boundaries, some nodes of the computational mesh might not belong to any patch associated to the background mesh and, consequently, the approach described above would ignore the metrics defined at those points. To avoid this, the nodes of the computational mesh that do not belong to any element of the background mesh are associated to the closest elements of the background mesh. This ensures that the all the metrics available in the computational mesh are utilised when transferring information to the background mesh.
\end{remark}

%============================================================================================
\subsection{ANN architecture to predict anisotropic spacing} \label{sc:spacingPrediction}
%============================================================================================

Using the strategies described in the two previous sections, the tensor metric that defines the target anisotropic spacing in a common background mesh is available, for a set of training cases.

A potential ANN architecture would involve setting, as inputs of the ANN, the desired parameters (e.g. flow conditions or geometric parameters) and defining, as outputs, the nine components of the metric tensor at all the nodes of the background mesh. However, such a strategy would ignore that a metric tensor must be given by a symmetric definite positive matrix~\cite{}. In addition, predicting the components of the metric tensor directly lacks some interpretability of the predictions as the information about the three mutually orthogonal directions and the spacing in each direction cannot be easily observed by looking at the components of the metric tensor directly.

The orthogonality property of the directions to be predicted can be exploited to reduce the amount of information to be predicted and at the same time ensure that the resulting metric tensor satisfies the required properties. To this end, the first direction (corresponding to the minimum spacing) is expressed using spherical coordinates as, meaning that only two angles are required, namely $\alpha_1$ and $\alpha_2$. Next, to strongly enforce the required orthogonality, the second direction must lie within the orthogonal plane to the first direction, meaning that it can be expressed in polar coordinates in the orthogonal plane using a single angle, namely $\alpha_3$. Finally, the third direction is uniquely determined from the orthogonality property and there is no need to characterise or predict this direction.

\begin{remark} \label{rk:metricSignIdentity}
Given the definition of a metric tensor in Equation~\eqref{eq:metric}, it is easy to verify that the metric tensor given by three spacial directions $\gamma\be_1$, $\be_2$ and $\be_3$, with spacings $\delta_1$, $\delta_2$ and $\delta_3$, respectively, is identical if $\gamma=1$ or $\gamma=-1$. Obviously, the same applies if the change of sign is applied to any of the three directions.
\end{remark}

The observation in Remark~\ref{rk:metricSignIdentity} implies that the angles can be restricted to $\alpha_1 \in [0, \pi]$, $\alpha_2 \in [-\pi/2, \pi/2]$ and  $\alpha_3\in [0, \pi]$. This observation also implies that two dissimilar angles, for instance $\epsilon$ and $\pi-\epsilon$, for a small angle $\epsilon$, would produce almost identical metric tensors. It is therefore not advisable to define the angles as the outputs of the ANN as a small variation of one anisotropic variation can lead to a large variation of the angle, making the training more difficult and leading to a lower prediction accuracy. 

To avoid this issue, each angle is independently mapped to an imaginary circle in the two dimensional plane, namely $\bv_i = \left( \cos(2 \alpha_i), \sin(2 \alpha_i) \right)$, ensuring that two dissimilar angles corresponding to almost identical orientations lead to almost identical vectors $\bv_i$. 

Therefore, the proposed approach consists of building an ANN where the inputs are flow conditions or geometric parameters and the outputs are the unit vectors $\bv_i$ and the corresponding spacings $\delta_i$ for $i=1,\ldots,\nsd$.

Three models have been developed and compared in the numerical examples shown later. The first model consists of training two different ANNs, one ANN to predict the three spacings and a second ANN to predict the three vectors $\bv_i$ that uniquely define the first two directions of anisotropy. The second model consists of training three different ANNs, one ANN to predict the three spacings, a second ANN to predict the two vectors $\bv_1$ and $\bv_2$ that uniquely define the first direction of anisotropy and a third ANN to predict the vector $\bv_3$ that uniquely defines the second direction of anisotropy. The third model investigated involves training four different ANNs, one ANN to predict the three spacings and three more ANNs to independently predict the vectors $\bv_1$, $\bv_2$ and $\bv_3$. Table~\ref{tb:NNTypes} summarises the three models described.
\begin{table}[!tb]
	\begin{center}
		\begin{minipage}{\textwidth}
			\caption{Three models used for training ANNs to predict the anisotropic spacing on the background mesh.}
			\label{tb:NNTypes}
			\begin{tabular}{@{}cllll@{}}
				\toprule
				Model & ANN architecture \\
				\midrule			
				1 & ANN$_1$ ($\delta_1$, $\delta_2$, $\delta_3$) & ANN$_2$ ($\alpha_1$, $\alpha_2$, $\alpha_3$)  \\
				2 & ANN$_1$ ($\delta_1$, $\delta_2$, $\delta_3$) & ANN$_2$ ($\alpha_1$, $\alpha_2$) & ANN$_3$ ($\alpha_3$) \\
				3 & ANN$_1$ ($\delta_1$, $\delta_2$, $\delta_3$) & ANN$_2$ ($\alpha_1$) & ANN$_3$ ($\alpha_2$) & ANN$_4$ ($\alpha_3$) \\
				\bottomrule
			\end{tabular}
		\end{minipage}
	\end{center}
\end{table}

In terms of implementation, TensorFlow 2.7.0~\cite{abadi2016tensorflow} is used to construct and train the ANNs. To minimise the effect of the random initialisation of the ANN weights, each training experiment is repeated five times with different random seeds. In all the examples, each training is performed for a maximum of 5,000 epochs, with early stopping applied if no improvement in the objective function is observed for 100 consecutive epochs. The batch size used in all experiments is eight, which produced better performance compared to the default value of 32 in TensorFlow.

Finally, to evaluate the prediction accuracy of the trained ANNs, the classical statistical $R^2$ measure~\cite{glantz2001primer} is used, with results reported separately for each of the predicted outputs.

%%%%%%%%%%%%%%%%%%%%%%%%%%%%%%%%%%%%%%%%%%%%%%%%%%%%%%%%%%%%%%%%%%%%%%%%%%%%%%%%%%%%%%%%%%%%%%%%%%%%%%%
\section{Numerical examples}\label{sc:examples}
%%%%%%%%%%%%%%%%%%%%%%%%%%%%%%%%%%%%%%%%%%%%%%%%%%%%%%%%%%%%%%%%%%%%%%%%%%%%%%%%%%%%%%%%%%%%%%%%%%%%%%%

This section presents two numerical examples to assess the accuracy of the proposed strategy to predict the anisotropic spacing for unseen simulations. The first example involves a problem with a fixed wing geometry and variable flow conditions characterised by two parameters. The second example involves a more complex problem with fixed flow conditions and 11 geometric parameters that characterise the shape of a wing in a full aircraft configuration. 

All the data used in the examples corresponds to inviscid flow simulations performed with the FLITE system~\cite{sorensen2003b}, a well established vertex-centred finite volume solver.

%-------------------------------------------------------------------------------
\subsection{Anisotropic spacing predictions on the ONERA M6 wing at various inflow conditions}\label{sc:M6exFlow}
%-------------------------------------------------------------------------------

The first example considers the inviscid compressible flow past the ONERA M6 wing for varying flow conditions, characterised by the free-stream Mach number, $M_\infty$, and the angle of attack $\alpha$. The variation of the flow conditions is $M_\infty \in [0.6,0.95]$ and $\alpha \in [0^\circ,8^\circ]$, encompassing subsonic and transonic flow regimes and leading to a substantial variation of the spacing function required for each simulation.

\begin{figure}[!tb] 
\centering
\subfigure[$M_\infty=0.660, \alpha =1.91^\circ$]{\includegraphics[width=0.3\columnwidth]{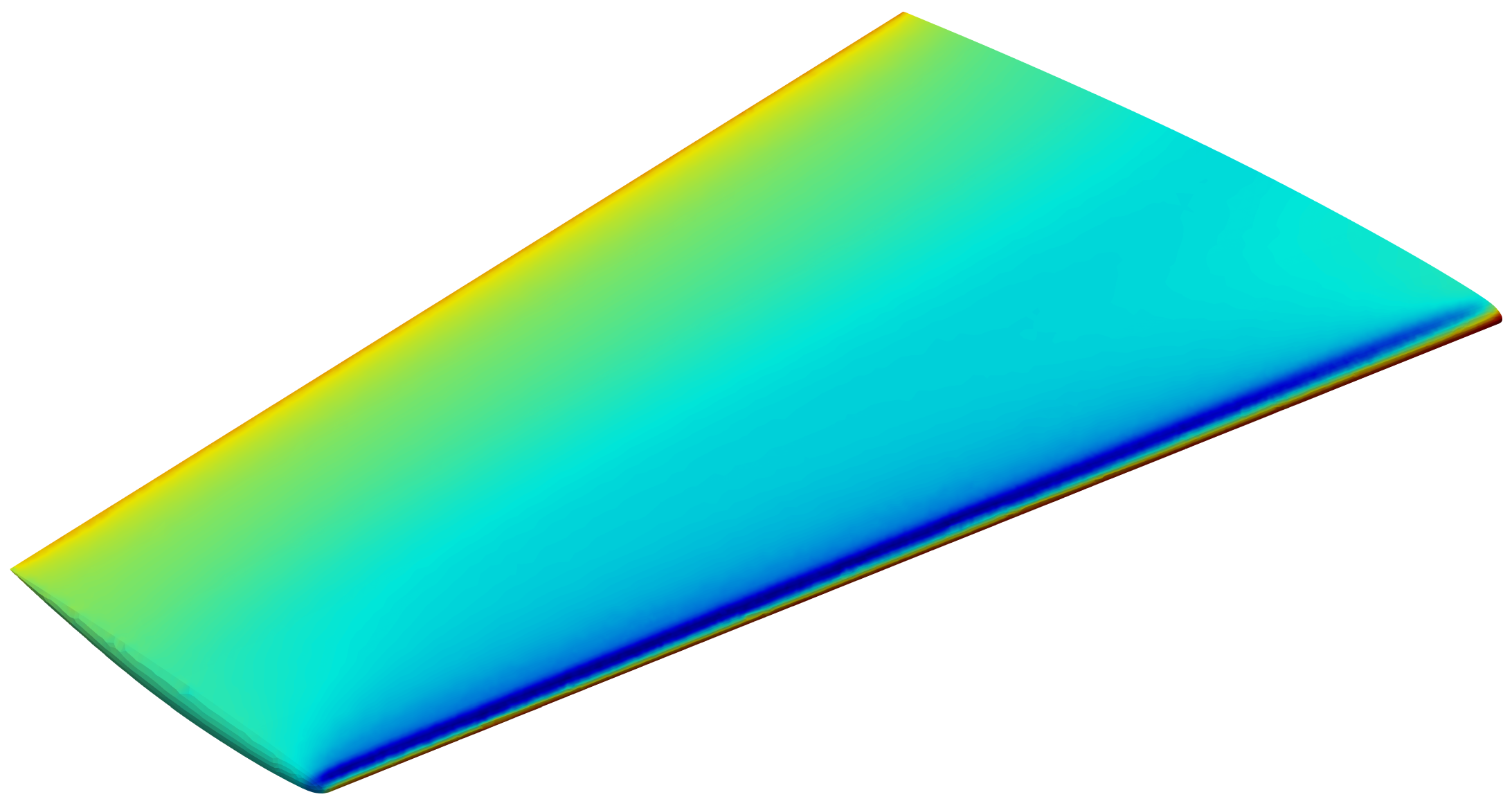}}
\subfigure[$M_\infty=0.885, \alpha =2.13^\circ$]{\includegraphics[width=0.3\columnwidth]{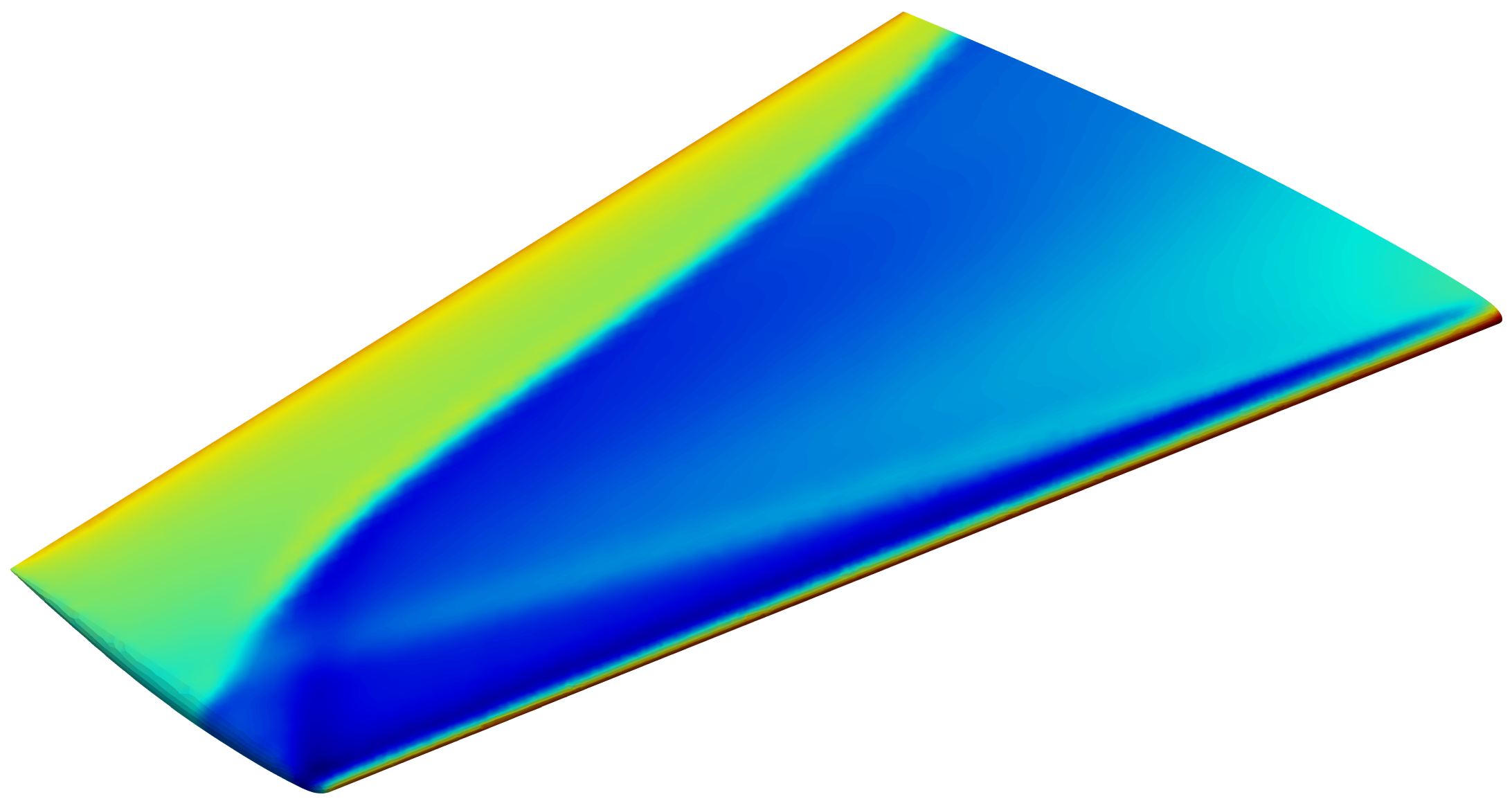}}
\subfigure[$M_\infty=0.809, \alpha =7.12^\circ$]{\includegraphics[width=0.3\columnwidth]{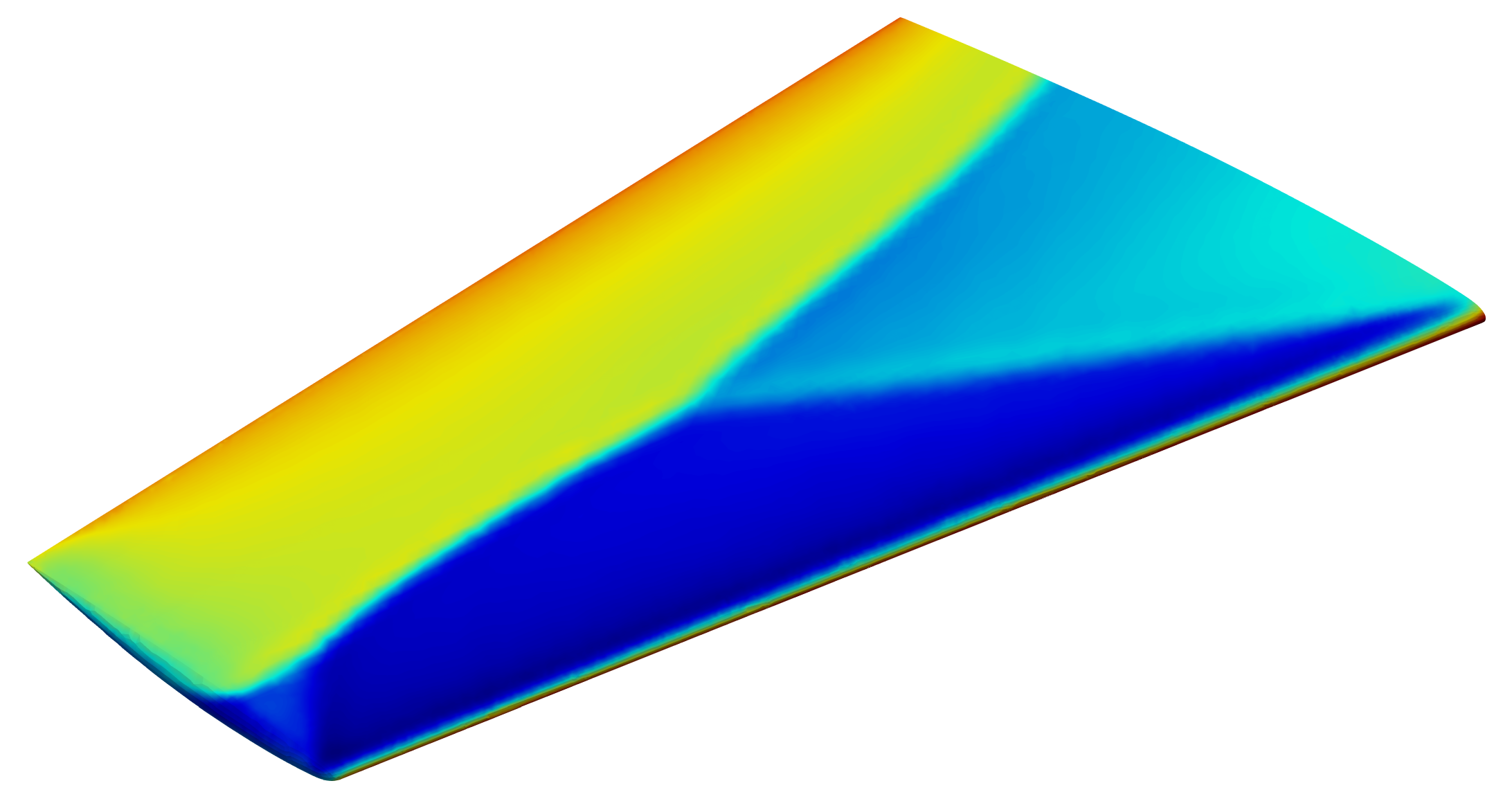}}
\caption{ONERA M6 wing: Pressure coefficient, $C_p$, for three different flow conditions.}
\label{fig:M6targetSols} 
\end{figure}

Figure~\ref{fig:M6targetSols} shows three different pressure coefficient distributions over the wing, corresponding to different flow conditions. 
The first case corresponds to a subsonic flow where the gradients are mainly concentrated along the leading and trailing edges and anisotropic spacing would be beneficial in these regions. The other two cases involve a transonic flow with a $\lambda$-shape shock, but with significant variability of the regions containing steep gradients. 

All the simulations available, including training and test cases, were computed on the same tetrahedral mesh with 4.6M elements and 782K nodes and using isotropic spacing. To select the training and test cases, sampling is performed using Halton sequencing~\cite{halton1964algorithm}. More precisely, scrambled Halton sequencing is employed as it is known to maintain the low-discrepancy in high-dimensional problems~\cite{ScrambledHaltonSequences}.

\begin{remark} \label{rk:sampling}
When utilising historical data available in industry, the datasets are unlikely to align with a Halton sequencing. Instead, it is anticipated that the sampling, done by an expert engineer, would be denser in critical regions of the flight envelope where significant changes in flow features are expected to occur. Therefore, although the influence of the sampling method is out of the scope of the current work, it is anticipated that a biased sampling, as done by an expert engineer, would lead to better performance of the trained ANN for the same amount of training data or to similar performance with less training data.
\end{remark}

For each available case, the metric tensor is computed at each node of the computational mesh using the procedure described in Section~\ref{sc:targetSpacingComp}. Next, employing the strategy presented in Section~\ref{sc:transferSpacing}, for each training case, the metric is transferred to the same background mesh, which in this example has approximately 590K elements and 100K nodes. Finally, using the procedure described in Section~\ref{sc:spacingPrediction}, the spacing and the vectors that describe the first two anisotropic directions are computed, for each case and for each node of the background mesh. 

\begin{figure}[!tb] 
	\centering
	\subfigure[$\delta_1$]{\includegraphics[width=0.32\textwidth]{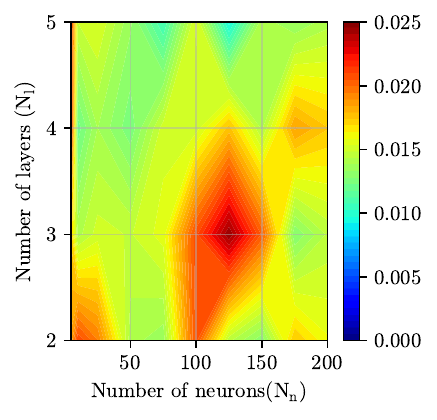}}
	\subfigure[$\delta_2$]{\includegraphics[width=0.32\textwidth]{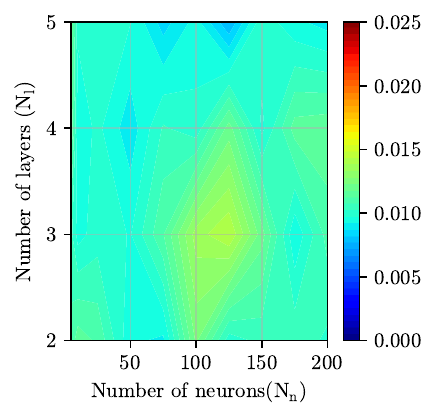}}
	\subfigure[$\delta_3$]{\includegraphics[width=0.32\textwidth]{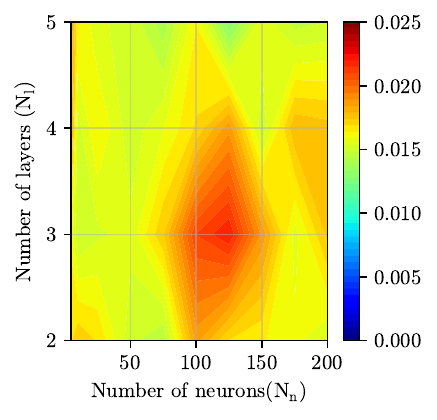}}
	\subfigure[$\alpha_1$]{\includegraphics[width=0.32\textwidth]{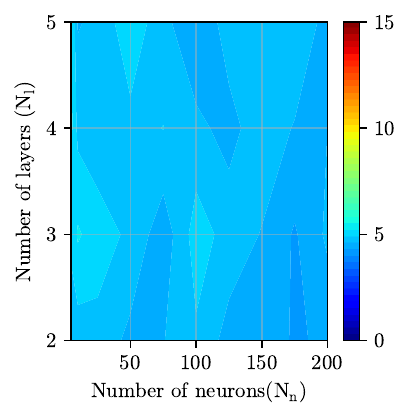}}
	\subfigure[$\alpha_2$]{\includegraphics[width=0.32\textwidth]{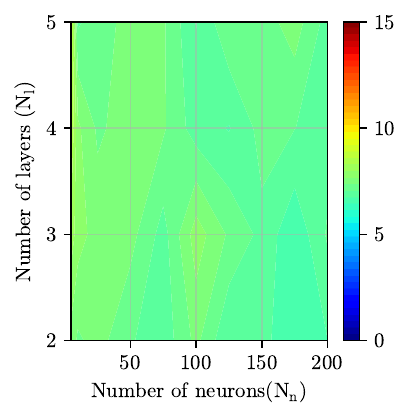}}
	\subfigure[$\alpha_3$]{\includegraphics[width=0.32\textwidth]{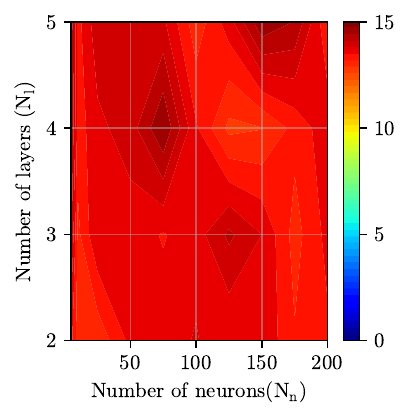}}
	\caption{ONERA M6 wing: MAE for the spacings ($\delta_i$) and angles ($\alpha_i$) as a function of the number of layers and number of neurons in each layer employing the first model of Table~\ref{tb:NNTypes}.}
	\label{fig:M6:Hyperparameters} 
\end{figure}

With this information ANNs are trained using the parameters described in Section~\ref{sc:spacingPrediction}. For the first model, described in Table~\ref{tb:NNTypes}, where two ANNs are trained separately to predict spacing and anisotropy directions, Figure~\ref{fig:M6:Hyperparameters}  shows the mean average error (MAE) for ANNs with an increasing number of hidden layers and neurons, when using $N_{tr} = 40$.
The results clearly show that predicting the spacing is significantly easier than predicting the anisotropic directions accurately, with the MAE for spacings being two or three orders of magnitude lower than that for the directions. The number of hidden layers utilised varies from two to five because previous studies involving the prediction of isotropic spacing~\cite{lock2023predicting} showed that one hidden layer produces significantly less accurate results and six layers do not bring any benefit in terms of accuracy. In fact, as shown in Figure~\ref{fig:M6:Hyperparameters}  using more than two layers does not provide any extra accuracy. The results also show that the most accurate results are obtained for the first anisotropic direction and its associated spacing. Given that the first direction is associated to the smaller spacing, this is the most important direction to be accurately predicted in order to capture the steep gradients near shocks. 

The number of neurons is varied between five and 200, but the results show that using more than 50 neurons does not provide any significant benefit. This means that the training times of the ANN are low. For instance using $N_{tr}=40$ two hidden layers and five neurons per layer, the training of the ANN to predict spacing with  takes approximately seven minutes, whereas the ANN to predict the anisotropic directions takes 35 minutes. When the number of neurons is increased to 100 per layer, the training of the ANN to predict spacing with  takes approximately 23 minutes, whereas the ANN to predict the anisotropic directions takes 67 minutes. 

\begin{figure}[!tb] 
	\centering
	\subfigure{\includegraphics[width=0.42\textwidth]{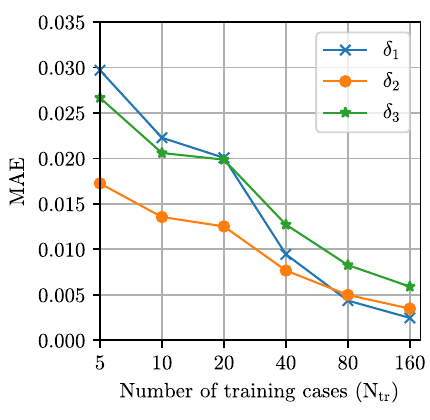}}
	\subfigure{\includegraphics[width=0.42\textwidth]{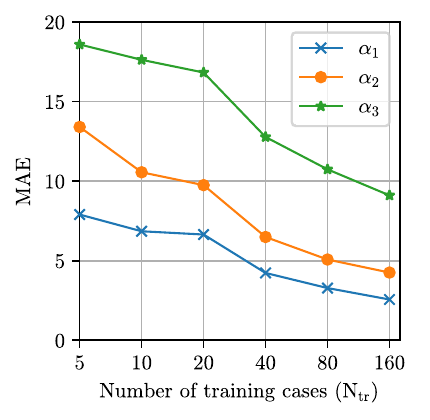}}
	\caption{ONERA M6 wing: MAE for the spacings ($\delta_i$) and angles ($\alpha_i$) as a function of the number of training cases, $N_{tr}$, employing the first model of Table~\ref{tb:NNTypes}.}
	\label{fig:M6:MAE_N_tr} 
\end{figure}

Next, the effect of the size of the training dataset is studied. For the first model described in Table~\ref{tb:NNTypes}, where two independent ANNs are trained, Figure~\ref{fig:M6:MAE_N_tr} shows the maximum MAE for all test cases as a function of the number of training cases for the six predicted outputs.
The ANN architecture used in this study considers the hyperparameters from the best performing ANN, as identified in the hyperparameter tuning process shown in Figure~\ref{fig:M6:Hyperparameters}. The results show a monotonous decrease of the MAE as the number of training cases is increased. For 40 training cases, a significant decrease in the MAE is observed for all the predicted outputs, suggesting that this number of cases provides enough information for the ANN to accurately predict the information required to produce suitable anisotropic spacings.

\begin{figure}[!tb] 
	\centering
	\subfigure[$M_\infty=0.66, \alpha =1.91^\circ$]{\includegraphics[width=0.32\textwidth]{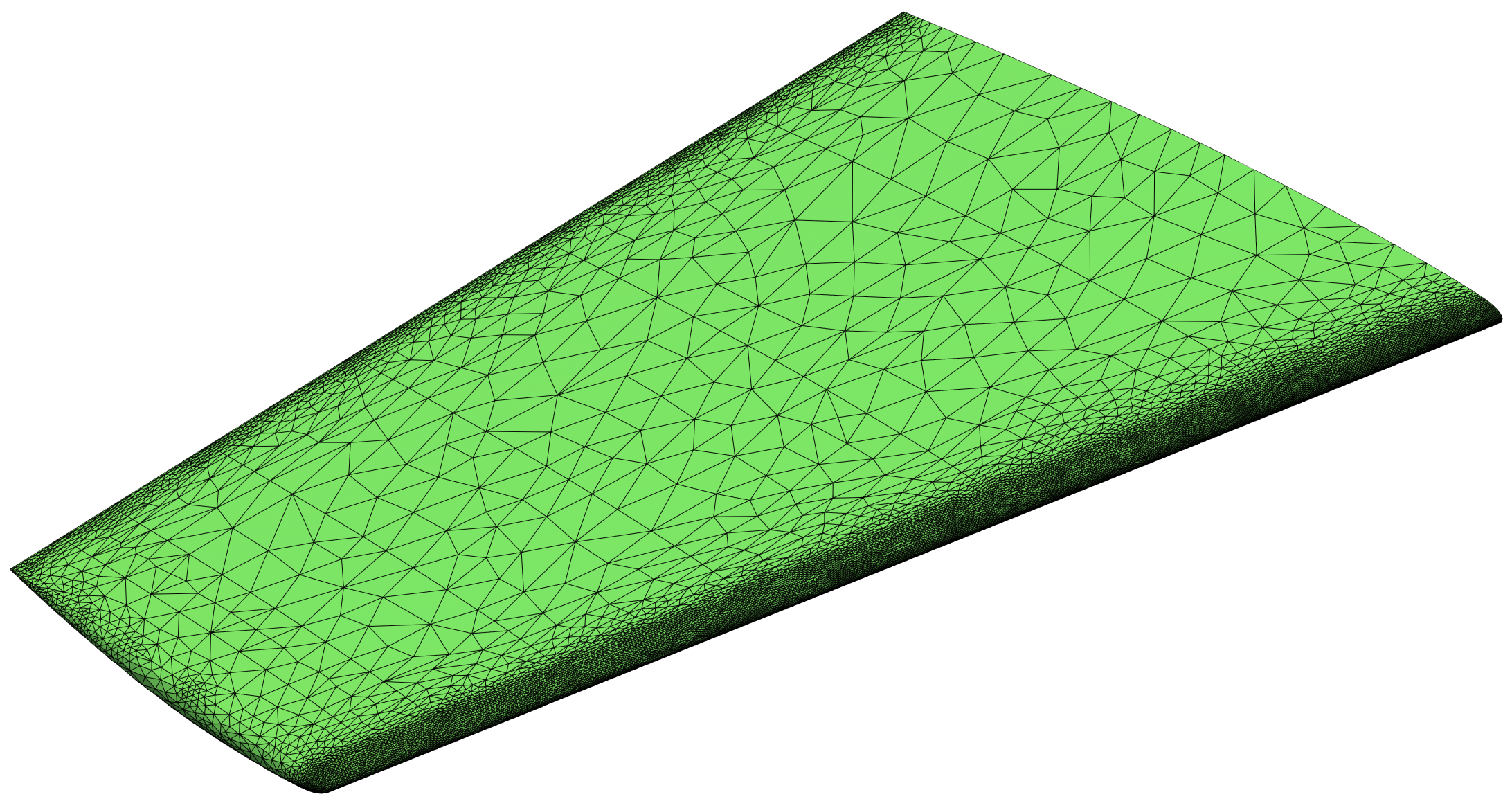}}
	\subfigure[$M_\infty=0.88, \alpha =2.13^\circ$]{\includegraphics[width=0.32\textwidth]{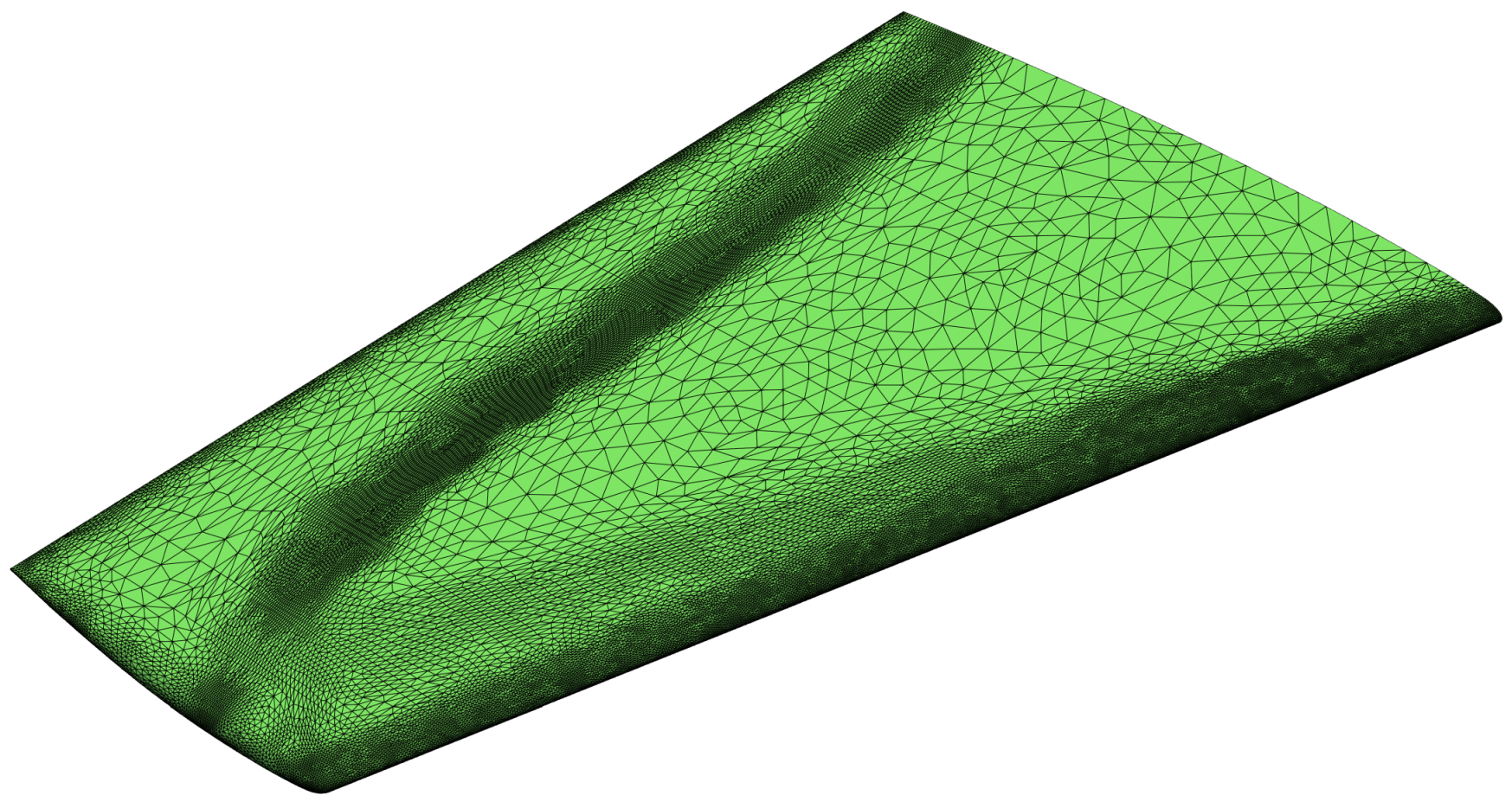}}
	\subfigure[$M_\infty=0.89, \alpha =7.12^\circ$]{\includegraphics[width=0.32\textwidth]{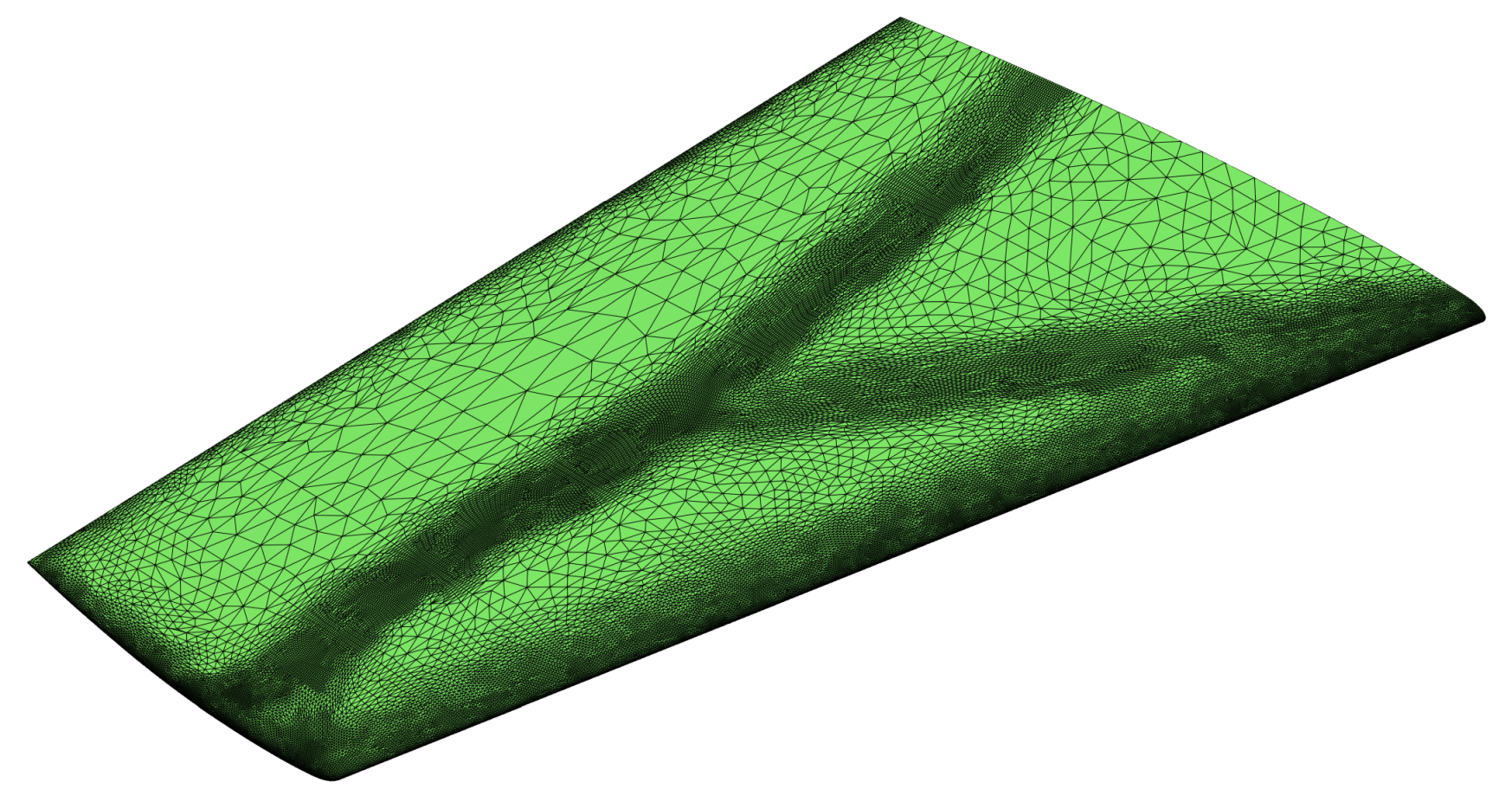}}
	\\
	\subfigure[$M_\infty=0.66, \alpha =1.91^\circ$]{\includegraphics[width=0.32\textwidth]{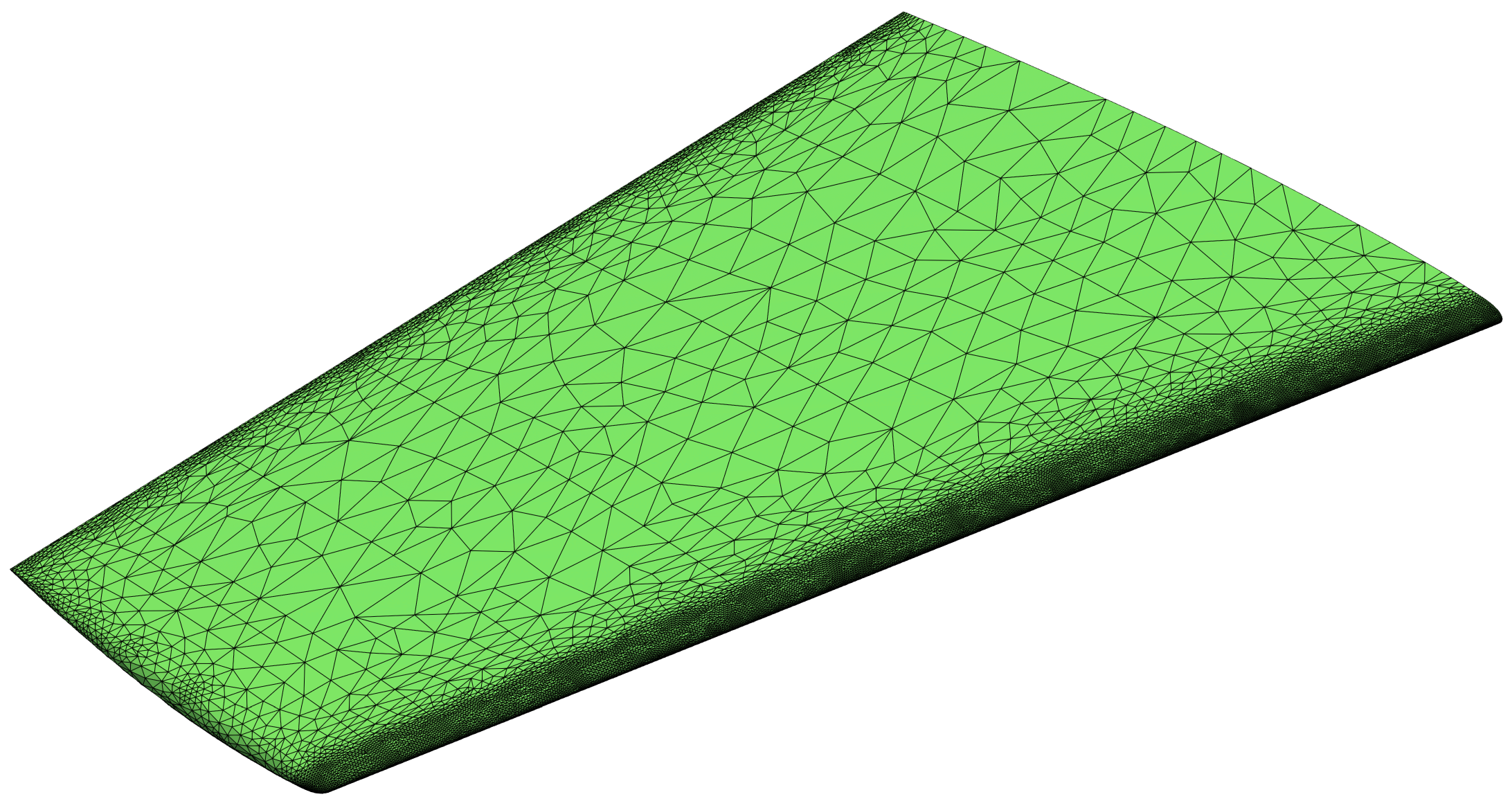}}
	\subfigure[$M_\infty=0.88, \alpha =2.13^\circ$]{\includegraphics[width=0.32\textwidth]{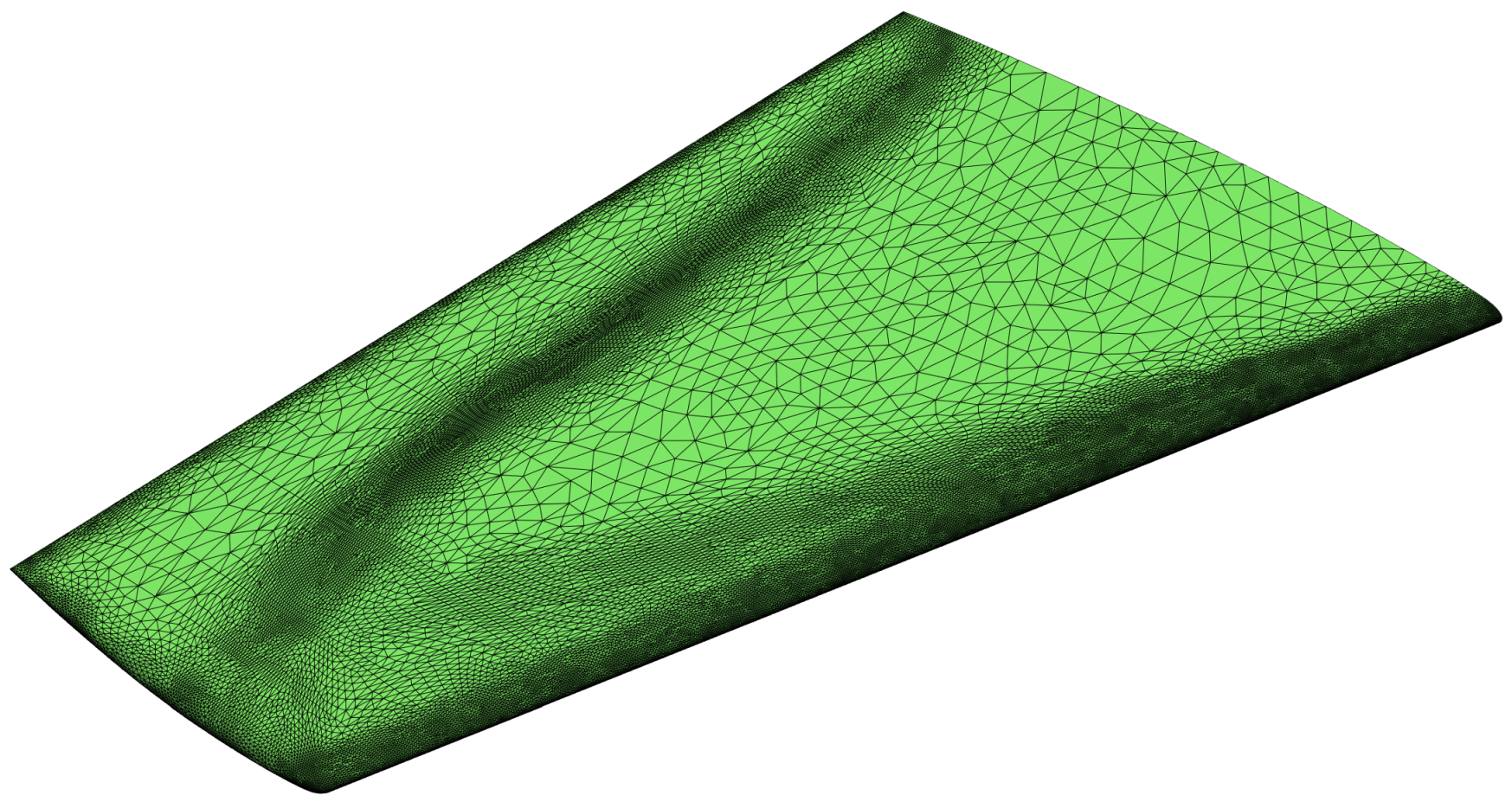}}
	\subfigure[$M_\infty=0.89, \alpha =7.12^\circ$]{\includegraphics[width=0.32\textwidth]{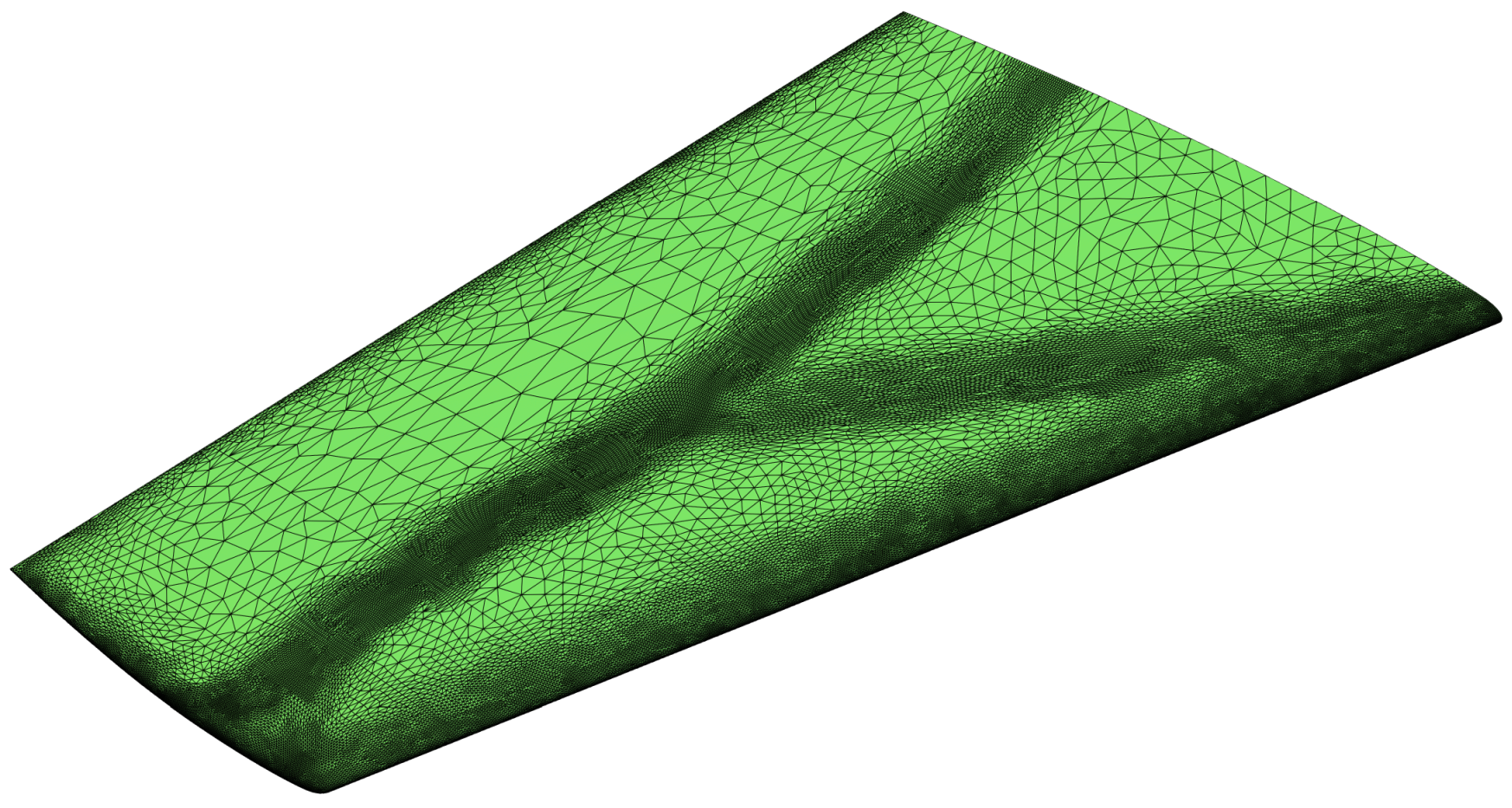}}
	\caption{ONERA M6 wing: Target (top) and predicted (bottom) meshes using the first model for three flow conditions unseen by the ANN during training.}
	\label{fig:M6:flowTargetPredictionMesh} 
\end{figure}

\begin{figure}[!tb] 
	\centering
	\subfigure[$M_\infty=0.88, \alpha =2.13^\circ$]{\includegraphics[width=0.49\textwidth]{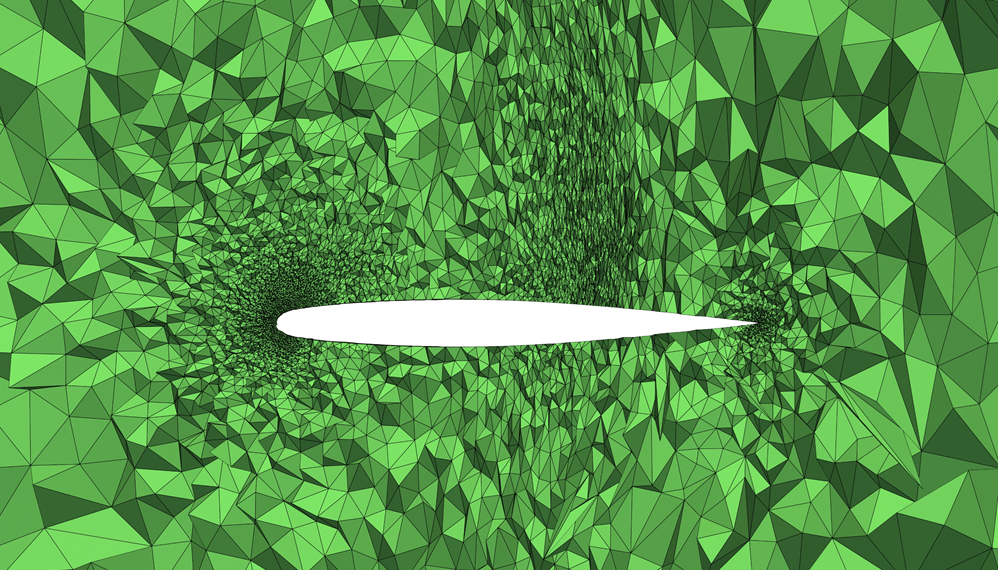}}
	\subfigure[$M_\infty=0.89, \alpha =7.12^\circ$]{\includegraphics[width=0.49\textwidth]{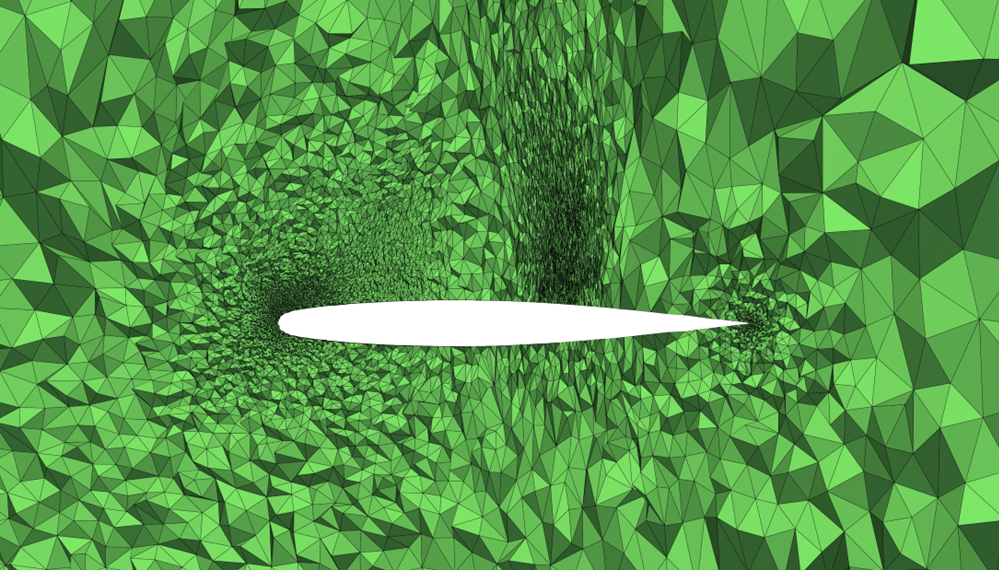}}
	\caption{ONERA M6 wing: predicted volume meshes using the first model for two transonic flow conditions unseen by the ANN during training.}
	\label{fig:M6:flowTargetPredictionMeshVolume} 
\end{figure}

To visually illustrate the potential of the proposed approach, the trained ANNs for the first model of Table~\ref{tb:NNTypes} are next used to predict the anisotropic spacing field, which is subsequently used to generate an anisotropic mesh. Figure~\ref{fig:M6:flowTargetPredictionMesh} shows the target mesh and the predictions for three inflow conditions corresponding to test cases, unseen by the ANN during training. 
The target meshes are generated by directly using the metric defined in the background mesh.
The predicted meshes closely follow the refinement patterns of the target meshes, effectively capturing key features of the flow. For the first case, which is a subsonic test case, the refinement is concentrated in the leading and trailing edges and the anisotropy of the predicted spacing matches the one observed in the target mesh. For the two transonic cases, the predicted meshes provide the refinement required to capture the shocks and, again, the anisotropic character of the target spacing is clearly observed in the predicted meshes. The anisotropic spacing is better illustrated for the two transonic cases by showing the volume mesh in Figure~\ref{fig:M6:flowTargetPredictionMeshVolume} .

To quantify the performance of the proposed strategy and to compare the different models in Table~\ref{tb:NNTypes}, the spacing function predicted by the ANN is compared with the target spacing function. It is worth noting that the spacings cannot be directly compared as the target and predicted directions of anisotropy are not identical. Therefore, to produce a suitable error measure, the predicted anisotropic spacings are projected onto the target spacing directions given by $\be_1$, $\be_2$ and $\be_3$ for each mesh node.

\begin{figure}[!tb] 
	\centering
	\subfigure[Spacing along $\textbf{e}_1$]{\includegraphics[width=0.49\textwidth]{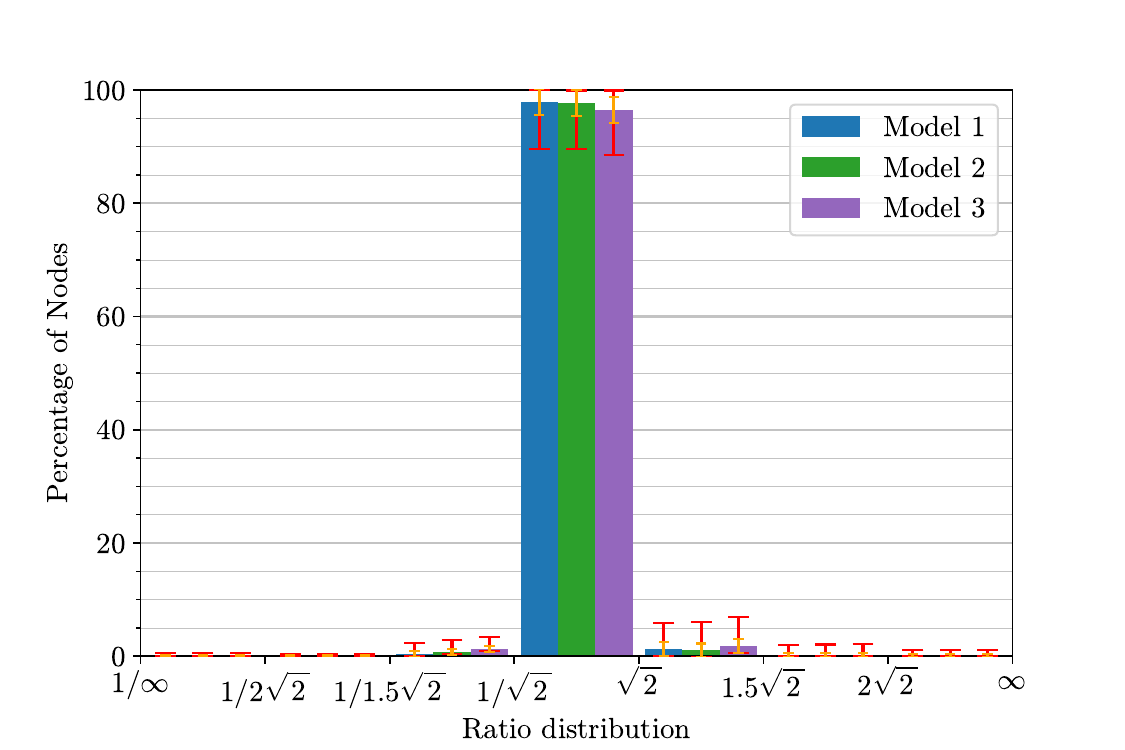}}
	\subfigure[Spacing along $\textbf{e}_2$]{\includegraphics[width=0.49\textwidth]{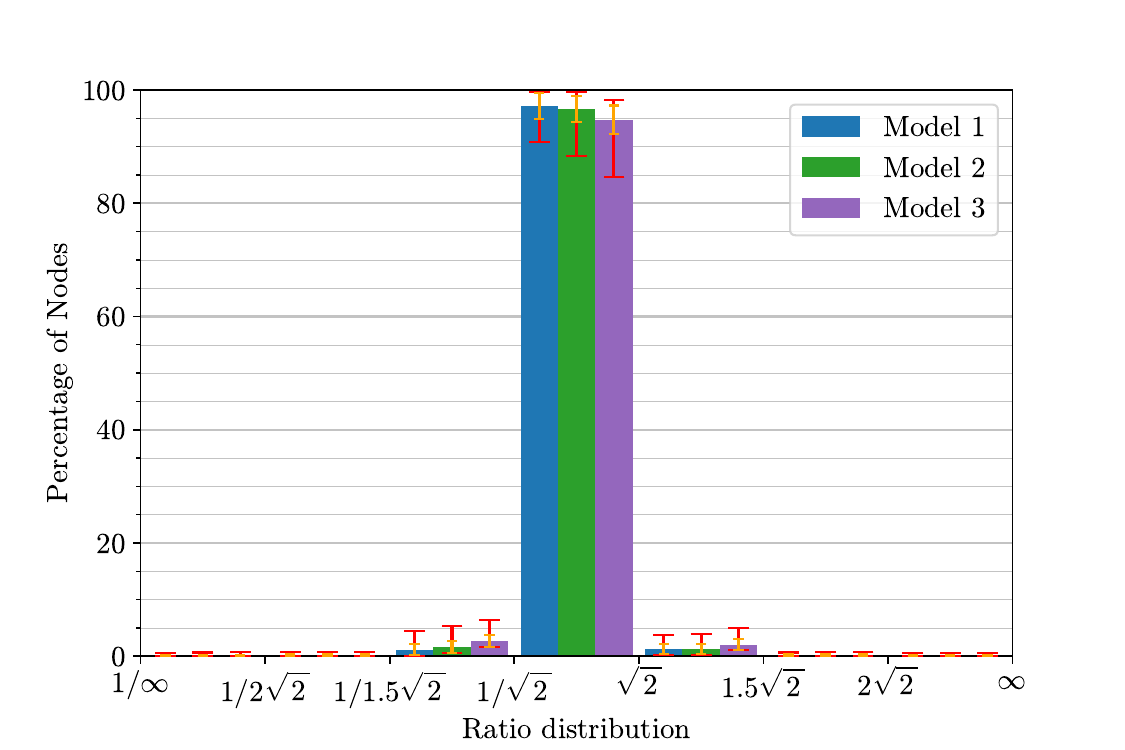}}
	\subfigure[Spacing along $\textbf{e}_3$]{\includegraphics[width=0.49\textwidth]{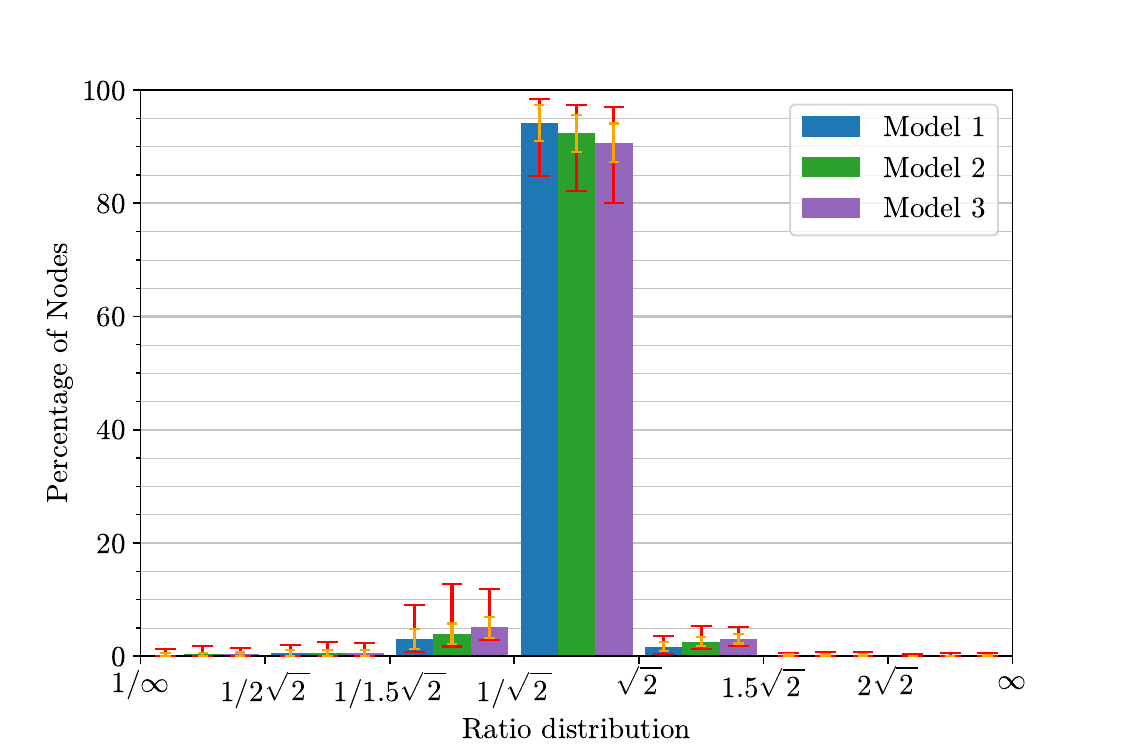}}
	\caption{ONERA M6 wing: Histogram of the ratio between the predicted and target spacings for the three models of Table~\ref{tb:NNTypes}.}
	\label{fig:M6:ModelSpacingComparison} 
\end{figure}

Figure~\ref{fig:M6:ModelSpacingComparison} presents histograms comparing the three models of Table~\ref{tb:NNTypes}, evaluated with $N_{tr} = 40$. 
The histograms show the mean ratio between predicted and target spacing along $\be_1$, $\be_2$, and $\be_3$, across all test cases. The red error bars indicate the minimum and maximum values for each bin, while the orange bars represent the standard deviation from the mean. The bins are defined as multiples of $\sqrt{2}$ and its inverse because, as usual in mesh generation, if a user-specified spacing $h$ at a point is sought, the acceptable values for the spacing are within $h/\sqrt{2}$ and $h\sqrt{2}$. In the histograms, ratios below $1/\sqrt{2}$ indicate more refinement than required, whereas ratios above $\sqrt{2}$ indicate that the mesh is not refined enough.

The results show that model 1, with two ANNs to predict spacing and directions separately, leads to the most accurate predictions for all directions of the target anisotropic spacing. Slightly less accurate results are obtained for the second model, where the angles associated to the first and second anisotropic directions are predicted by two independent ANNs. Finally, the least accurate results are observed for the third model, where each angle is predicted by a different ANN. This indicates that using a single ANN to predict all the information about anisotropic directions is not only more efficient but also provides the best accurate predictions. 

Very small differences between the accuracy of the three models are observed on the predicted spacing along the first target anisotropic direction, $\be_1$, with almost 95\% of the mesh nodes having an acceptable prediction with the first model. For the second target anisotropic direction, $\be_2$, the third model shows a sizeable loss of accuracy, whereas the first and second models have almost the same accuracy. The better performance of the first model is best appreciated when considering the spacing along the third target anisotropic direction, $\be_3$. It is also worth mentioning that the first model is the one that exhibits the minimum areas of under-refinement.

The best performance of the first model is attributed to its ability to leverage global information during training as it is the only model that predicts the two first directions of anisotropy together.

\begin{figure}[!tb] 
	\centering
	\subfigure[Spacing in $\textbf{e}_1$]{\includegraphics[width=0.49\textwidth]{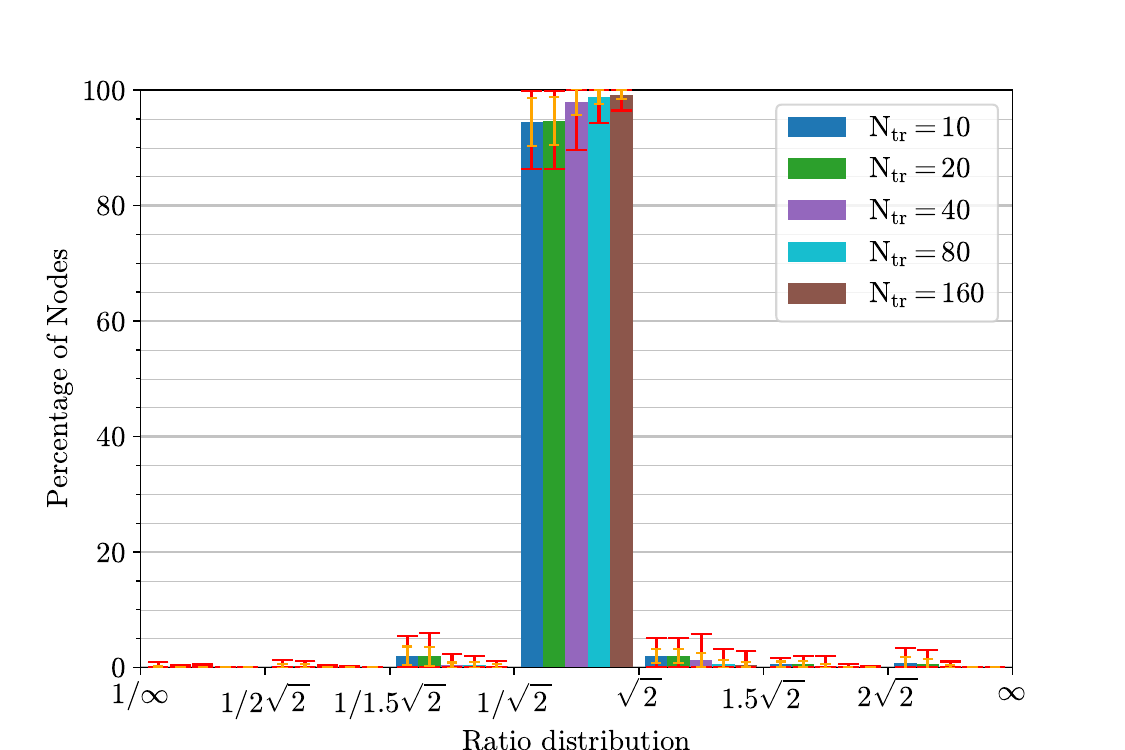}}
	\subfigure[Spacing in $\textbf{e}_2$]{\includegraphics[width=0.49\textwidth]{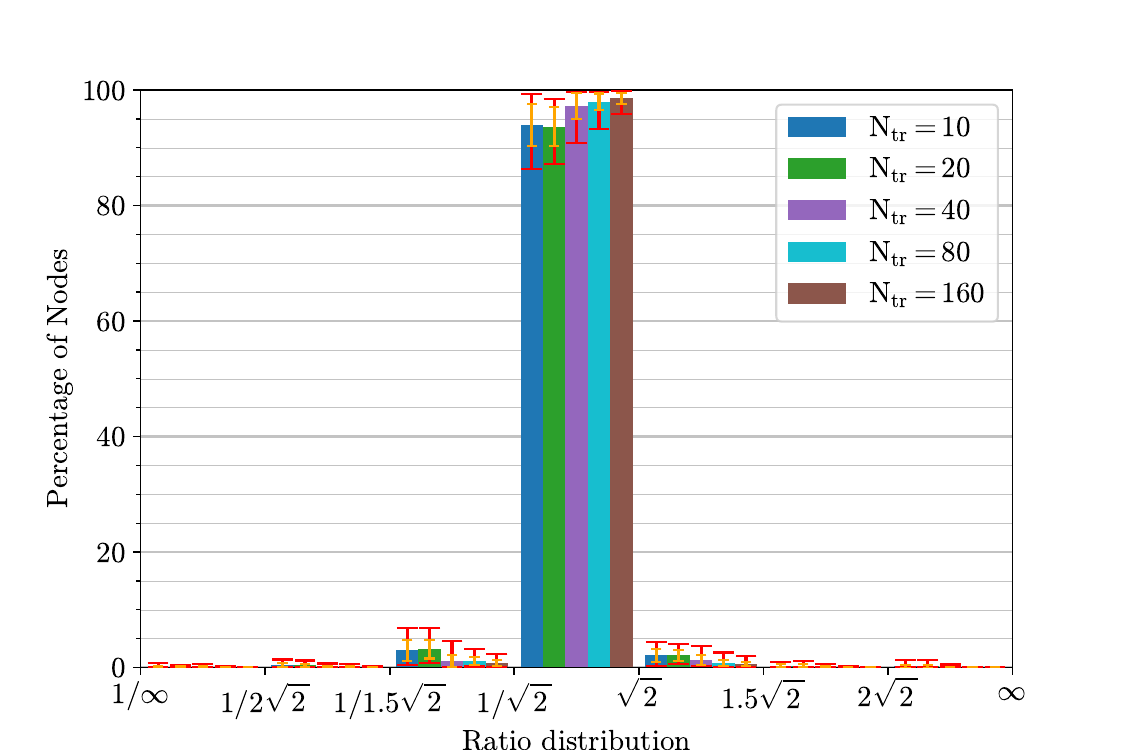}}
	\subfigure[Spacing in $\textbf{e}_3$]{\includegraphics[width=0.49\textwidth]{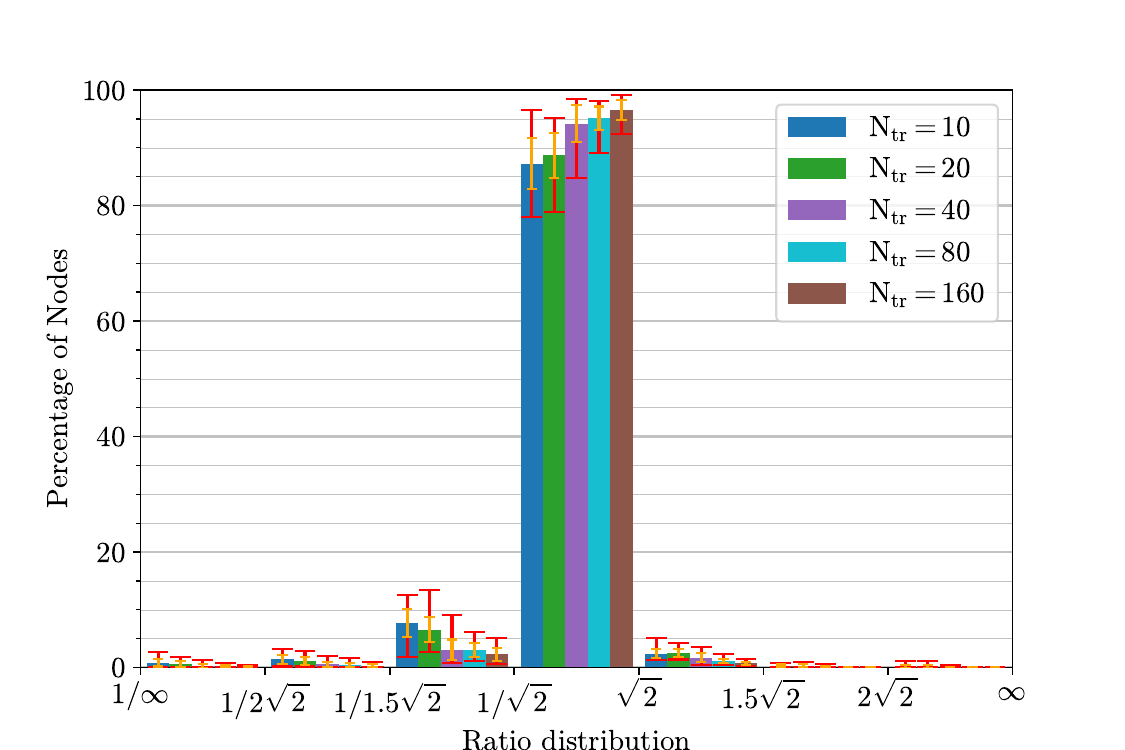}}
	\caption{ONERA M6 wing: Histogram of the ratio between the predicted and target spacings for the first model of Table~\ref{tb:NNTypes} and for an increasing number of training cases, $N_{tr}$.}
	\label{fig:M6NtrSpacingComparison} 
\end{figure}

To further quantify the accuracy of the predictions in terms of the training dataset size, Figure~\ref{fig:M6NtrSpacingComparison}  shows the histogram of the anisotropic spacing accuracy for the first model in terms of the number of training cases, $N_{tr}$.
The histograms show a significant improvement in prediction accuracy for $N_{tr} \geq 40$, as previously shown with the MAE graphs in Figure~\ref{fig:M6:MAE_N_tr}. It is worth noting that, for $N_{tr} = 40$ the percentage of nodes with a prediction of the full metric tensor reaches almost 95\%. Furthermore, with only 20 training cases, the prediction of the spacing in the first direction (the most critical to accurately represent the shocks) is acceptable for almost 95\% of the nodes.

\begin{figure}[!tb] 
	\centering
	\subfigure[$M_\infty=0.66, \alpha =1.91^\circ$]{\includegraphics[width=0.32\textwidth]{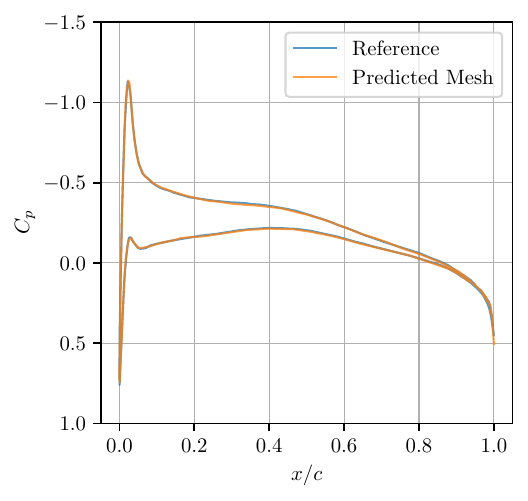}}
	\subfigure[$M_\infty=0.88, \alpha =2.13^\circ$]{\includegraphics[width=0.32\textwidth]{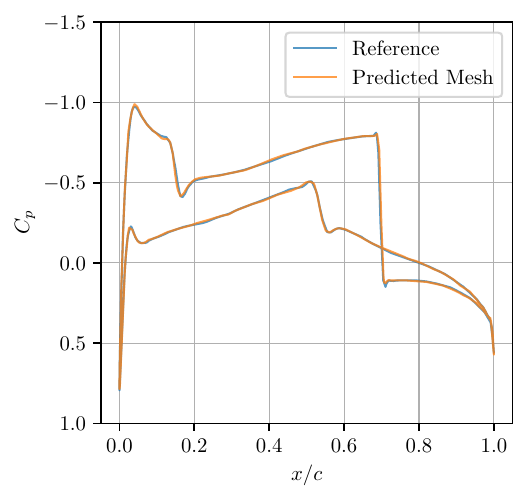}}
	\subfigure[$M_\infty=0.89, \alpha =7.12^\circ$]{\includegraphics[width=0.32\textwidth]{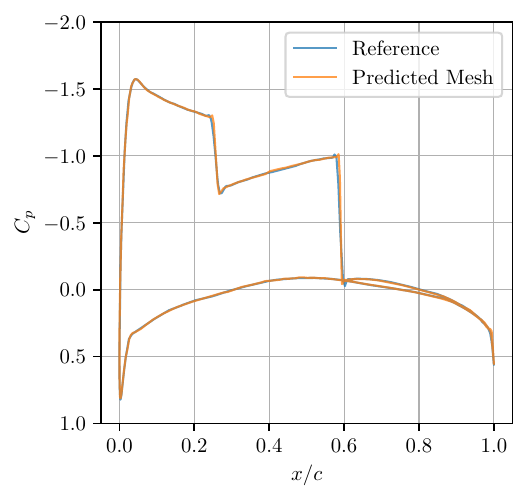}}
	\caption{ONERA M6 wing: Comparison of the pressure coefficient, $C_p$, for three different unseen flow conditions, at one section.}
	\label{fig:M6:flowTargetPredictionCp} 
\end{figure}

To conclude this example, the suitability of the meshes predicted to perform simulations for unseen cases is studied. To this end, an anisotropic spacing is predicted for three unseen cases, the corresponding anisotropic meshes are generated and simulations are performed. The results, involving pressure coefficient distribution and the lift and drag coefficients, are compared to the results obtained with a fine reference mesh which is produced using mesh adaptivity, starting with the original isotropic mesh with 4.6M elements used to generate all the datasets. Figure~\ref{fig:M6:flowTargetPredictionCp} shows the excellent agreement between the pressure coefficient distributions at one section of the wing for three different flow conditions unseen during the ANN training.

To confirm the suitability of the predicted meshes, Table~\ref{tb:M6aeroPrediction} compares the reference lift and drag coefficients, $C_L$ and $C_D$ respectively, with the one obtained after performing a simulation with the predicted meshes.
\begin{table}[!tb]
	\begin{center}
		\begin{minipage}{\textwidth}
			\caption{ONERA M6 wing: Reference aerodynamic coefficients and values computed with the predicted meshes.}
			\label{tb:M6aeroPrediction}
			\begin{tabular*}{\textwidth}{@{\extracolsep{\fill}}lcccccc@{\extracolsep{\fill}}}
				\toprule%
				& \multicolumn{2}{@{}c@{}}{$M_\infty=0.660, \alpha=1.91^\circ$} &
				\multicolumn{2}{@{}c@{}}{$M_\infty=0.885, \alpha=2.13^\circ$} &
				\multicolumn{2}{@{}c@{}}{$M_\infty=0.809, \alpha=7.12^\circ$}
				\\ \cmidrule{2-3}\cmidrule{4-5}\cmidrule{6-7}%
				& Reference & Predicted & Reference & Predicted & Reference & Predicted \\
				\midrule
				$C_{L}$ & 0.146   &  0.147   & 0.240   &  0.241   & 0.645   &  0.646 \\
				$C_{D}$ & 0.0014  &  0.0013  & 0.0142  &  0.0143  & 0.0628  &  0.0630 \\
				\bottomrule
			\end{tabular*}
		\end{minipage}
	\end{center}
\end{table}
The results show a maximum variation in the lift and drag coefficients of only one lift count and two drag counts respectively, clearly demonstrating the suitability of the predicted meshes to perform simulations of unseen cases.

%-------------------------------------------------------------------------------
\subsection{Anisotropic spacing predictions on a geometrically parametrised aircraft}\label{sc:falcon}
%-------------------------------------------------------------------------------

The second example considers the anisotropic spacing prediction for a full aircraft configuration that is geometrically parametrised at fixed transonic flow conditions corresponding to free-stream Mach number $M_\infty = 0.78$ and angle of attack $\alpha = 2.0^\circ$.

The wings of the aircraft are parametrised using 11 geometric parameters, with the details given in Table~\ref{tab:Falcon:design_parameters} and for a fixed total semi-span of 8.15m.
\begin{table}[!tb]
	\centering
	\caption{Geometrically parametrised aircraft: design parameters and the range of variation for each parameter.}
	\label{tab:Falcon:design_parameters}
	\begin{tabular}{lcc}
		\toprule
		\textbf{Parameter} & \textbf{Lower Limit} & \textbf{Upper Limit} \\
		\midrule
		Inboard sweep 				& $15^\circ$ & $50^\circ$ \\
		Outboard sweep 				& $15^\circ$ & $60^\circ$ \\
		Inboard semi-span 			& $2.0$ m & $5.0$ m \\
		Inboard dihedral			& $-7.5^\circ$ & $7.5^\circ$ \\
		Outboard dihedral 			& $-7.5^\circ$ & $7.5^\circ$ \\
		Mid-span twist 				& $0^\circ$ & $3^\circ$ \\
		Tip twist 					& $-2^\circ$ & $3^\circ$ \\
		Mid-span chord ratio 		& $0.25$ & $0.75$ \\
		Tip chord ratio 			& $0.25$ & $0.75$ \\
		Mid-span thickness ratio	& $0.25$ & $0.75$ \\
		Tip thickness ratio 		& $0.25$ & $0.75$ \\
		\bottomrule
	\end{tabular}
\end{table}

\begin{figure}[!tb] 
	\centering
	\subfigure[Geometry 1]{\includegraphics[width=0.32\columnwidth]{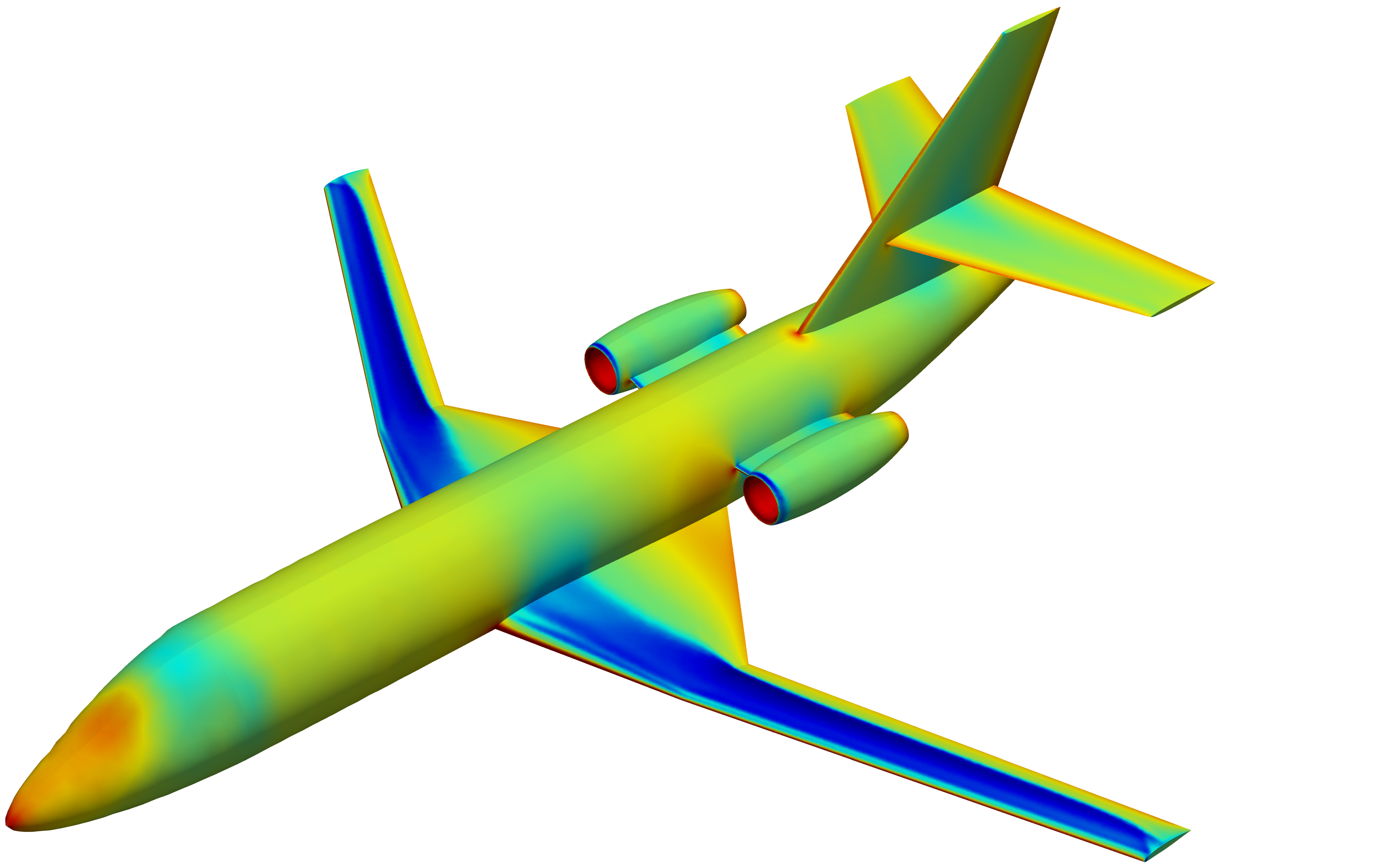}}
	\subfigure[Geometry 2]{\includegraphics[width=0.32\columnwidth]{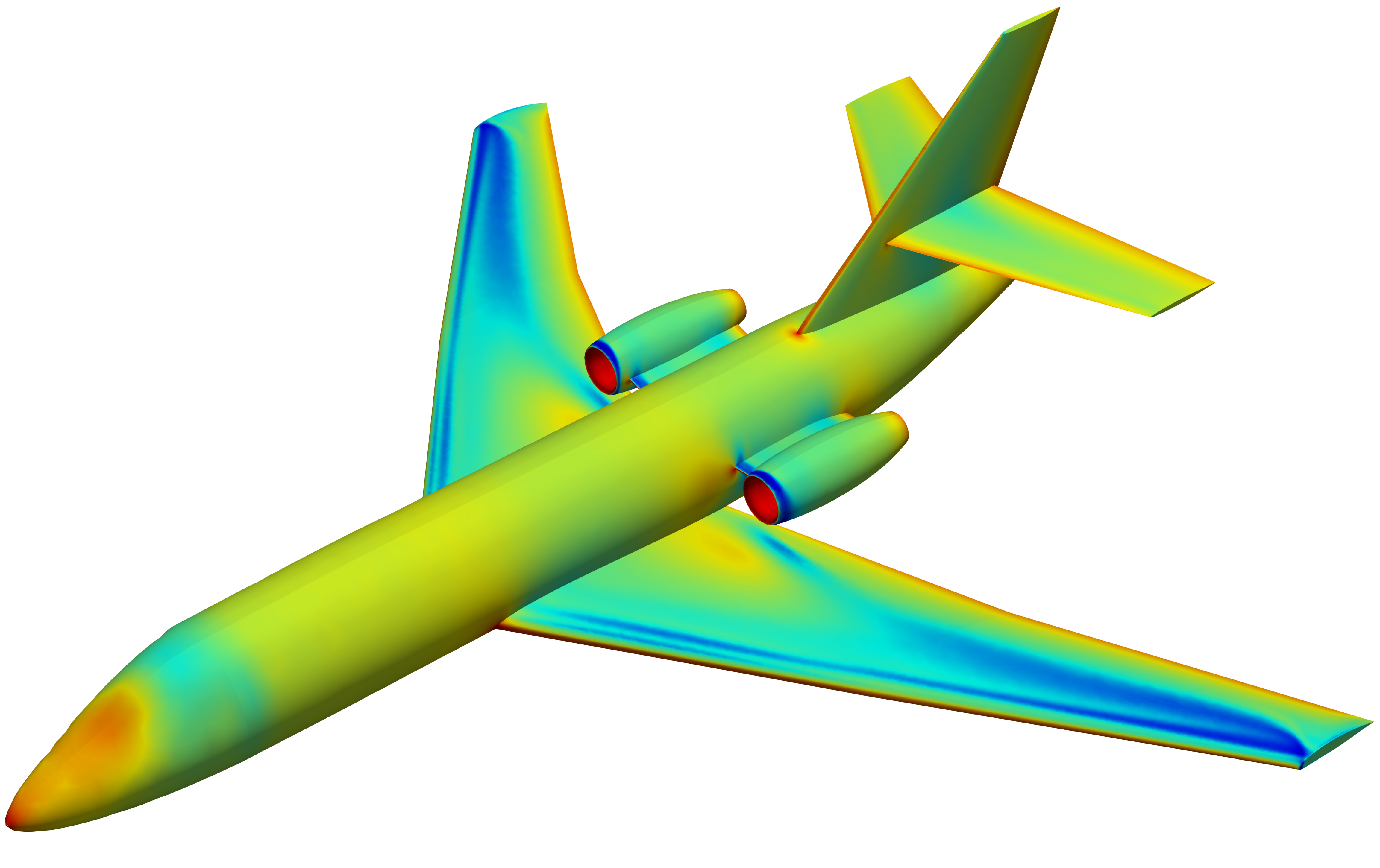}}
	\subfigure[Geometry 3]{\includegraphics[width=0.32\columnwidth]{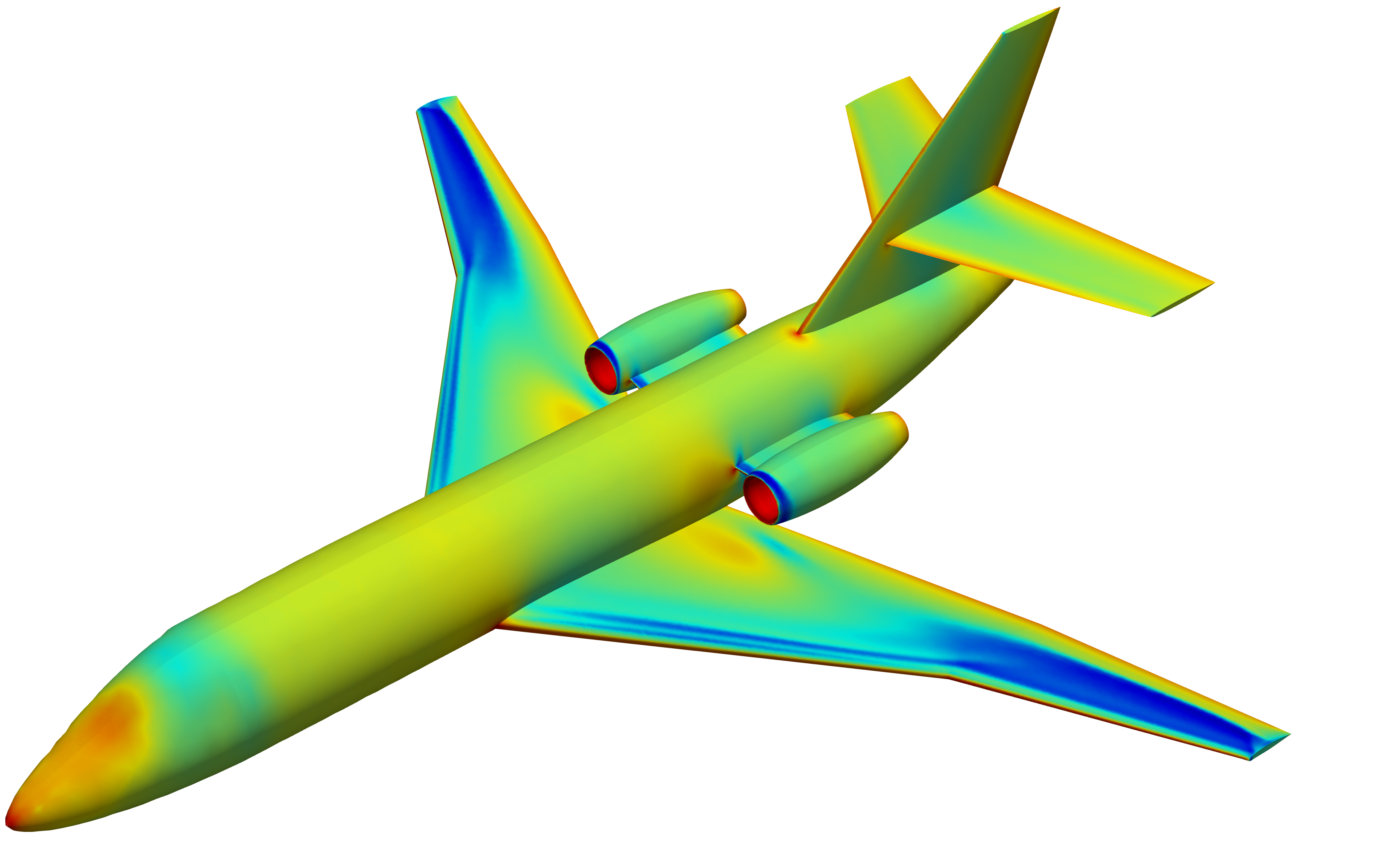}}
	\caption{Geometrically parametrised aircraft: Pressure coefficient, $C_p$, for three different geometric configurations.}
	\label{fig:Falcon:FlowEx} 
\end{figure}

To illustrate the variability in the geometry and the corresponding solution induced by the selected geometric parameters, Figure~\ref{fig:Falcon:FlowEx} depicts the pressure coefficient for three different geometric configurations.
In the three cases shown, a strong shock wave is observed on the upper surface of the wing, clearly displaying the transonic nature of the flow. With the first geometry, a well-defined shock wave is observed along the entire upper surface of the wing. With the second geometry, the shock structure changes significantly and the interaction of the wing and engine intake can be observed. Lastly, with the third geometry, a stronger shock is present over the outer section of the wing, and a complex interaction of the wing and engine is observed. These examples highlight the significant impact of the geometry on the flow features and emphasise the importance of using meshes tailored for each simulation.

As in the previous example, the strategy presented in the previous section is used to compute the metric tensor for each available simulation and to transfer the spacing to the background mesh, which in this example has approximately 847K elements and 143K nodes. 

\begin{figure}[!tb] 
	\centering
	\subfigure[$\delta_1$]{\includegraphics[width=0.32\textwidth]{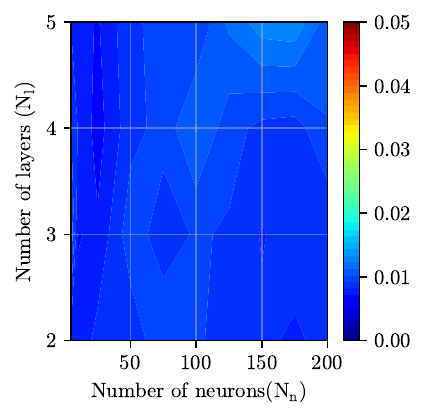}}
	\subfigure[$\delta_2$]{\includegraphics[width=0.32\textwidth]{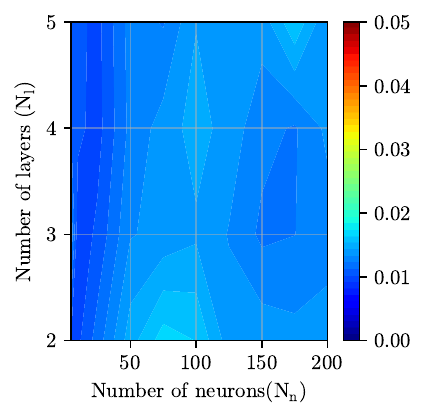}}
	\subfigure[$\delta_3$]{\includegraphics[width=0.32\textwidth]{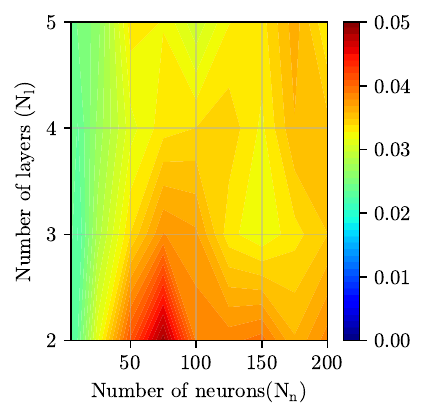}}
	\subfigure[$\alpha_1$]{\includegraphics[width=0.32\textwidth]{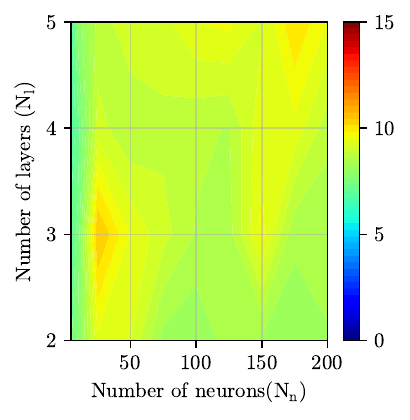}}
	\subfigure[$\alpha_2$]{\includegraphics[width=0.32\textwidth]{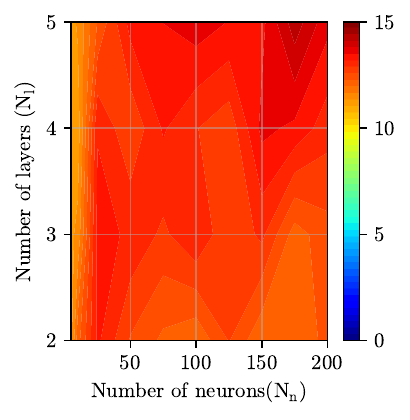}}
	\subfigure[$\alpha_3$]{\includegraphics[width=0.32\textwidth]{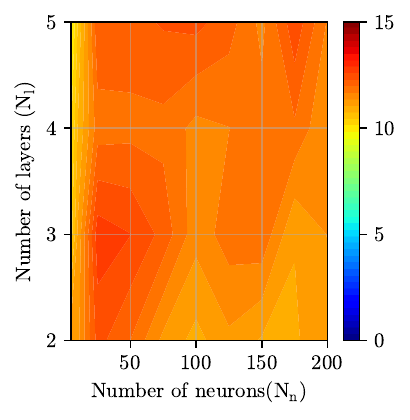}}
	\caption{Geometrically parametrised aircraft: MAE for the spacings ($\delta_i$) and angles ($\alpha_i$) as a function of the number of layers and number of neurons in each layer employing the first model of Table~\ref{tb:NNTypes}.}
	\label{fig:Falcon:Hyperparameters} 
\end{figure}

The ANNs are trained using the same parameters as in the previous example, as described in Section~\ref{sc:spacingPrediction}. Given the better performance and efficiency of the first model described in Table~\ref{tb:NNTypes}, this is the only model considered in this example, where two ANNs are trained separately to predict spacing and anisotropy directions. Figure~\ref{fig:Falcon:Hyperparameters}  shows the MAE for ANNs with an increasing number of hidden layers and neurons, when using $N_{tr} = 40$.
The results show a qualitative behaviour similar to the previous example despite the different nature of the parameters and the increased dimensionality of the problem. The prediction of the spacing is again significantly more accurate than the prediction of the angles that define the directions of anisotropy. The main difference with respect to the previous example is that the prediction of the angles that define the anisotropic direction is more challenging, in particular the second angle, $\alpha_2$. This is expected given the fact that the same amount of data is utilised, $N_{tr}=40$, but this problem has 11 geometric parameters.

\begin{figure}[!tb] 
	\centering
	\subfigure{\includegraphics[width=0.42\textwidth]{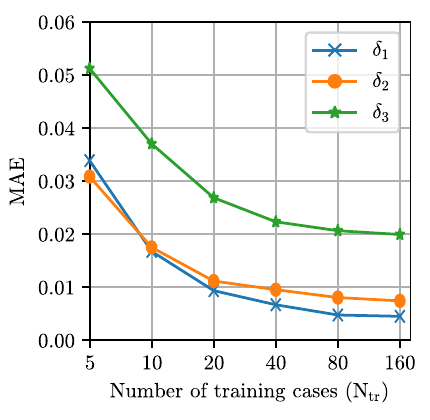}}
	\subfigure{\includegraphics[width=0.42\textwidth]{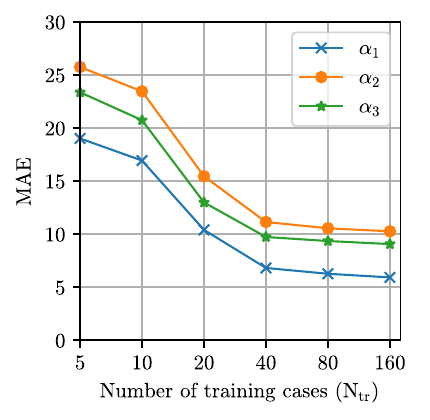}}
	\caption{Geometrically parametrised aircraft: MAE for the spacings ($\delta_i$) and angles ($\alpha_i$) as a function of the number of training cases, $N_{tr}$, employing the first model of Table~\ref{tb:NNTypes}.}
	\label{fig:Falcon:MAE_N_tr} 
\end{figure}

To study the effect of the training dataset size, Figure~\ref{fig:Falcon:MAE_N_tr} the maximum MAE for all test cases as a function of the number of training cases for the six predicted outputs.
As in the previous example the error decreases as the number of training cases increases. Despite the increased difficulty of this problem, it is remarkable to observe a significant gain in accuracy with only 40 training cases.

\begin{figure}[!tb] 
	\centering
	\subfigure[Geometry 1]{\includegraphics[width=0.32\textwidth]{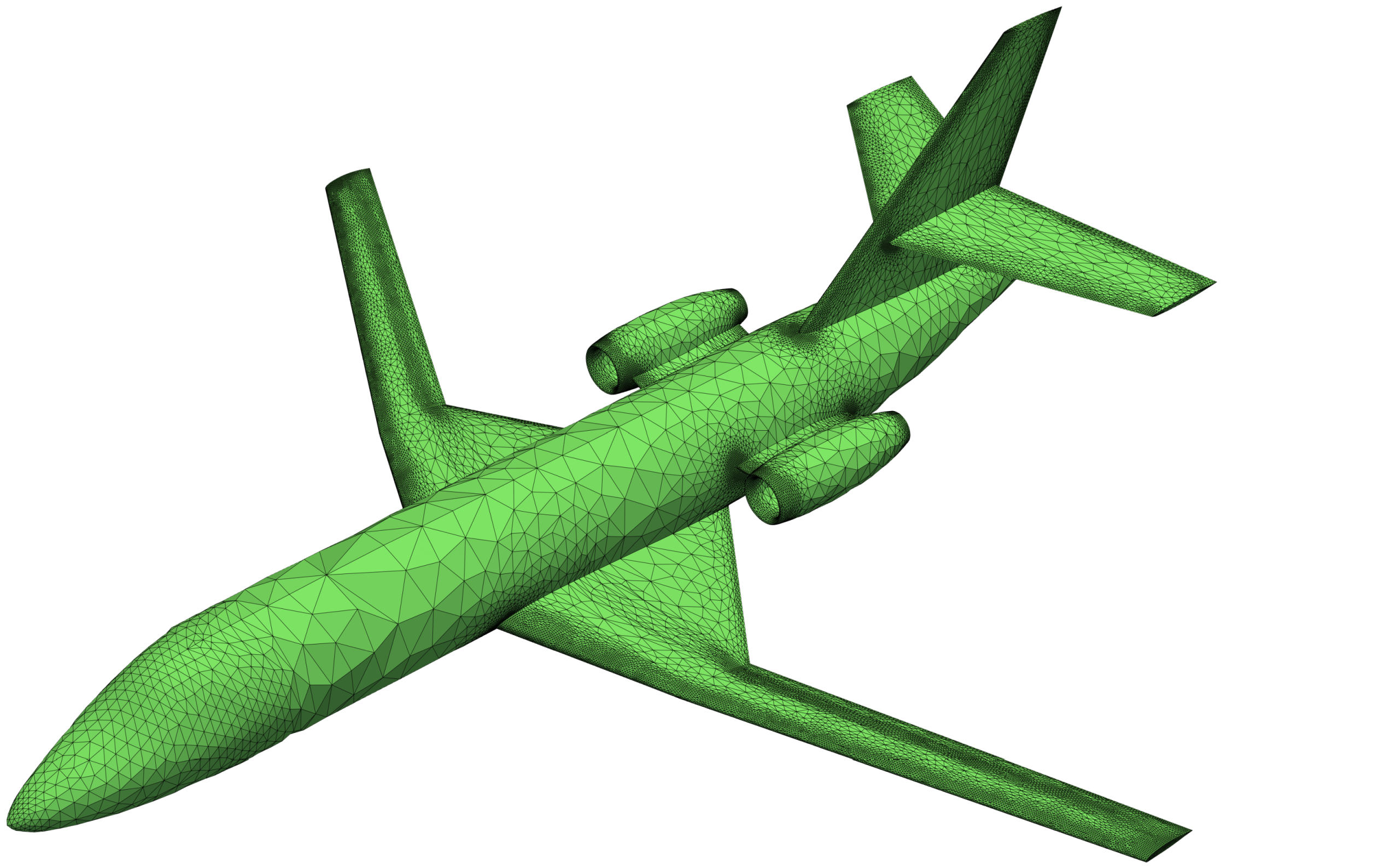}}
	\subfigure[Geometry 2]{\includegraphics[width=0.32\textwidth]{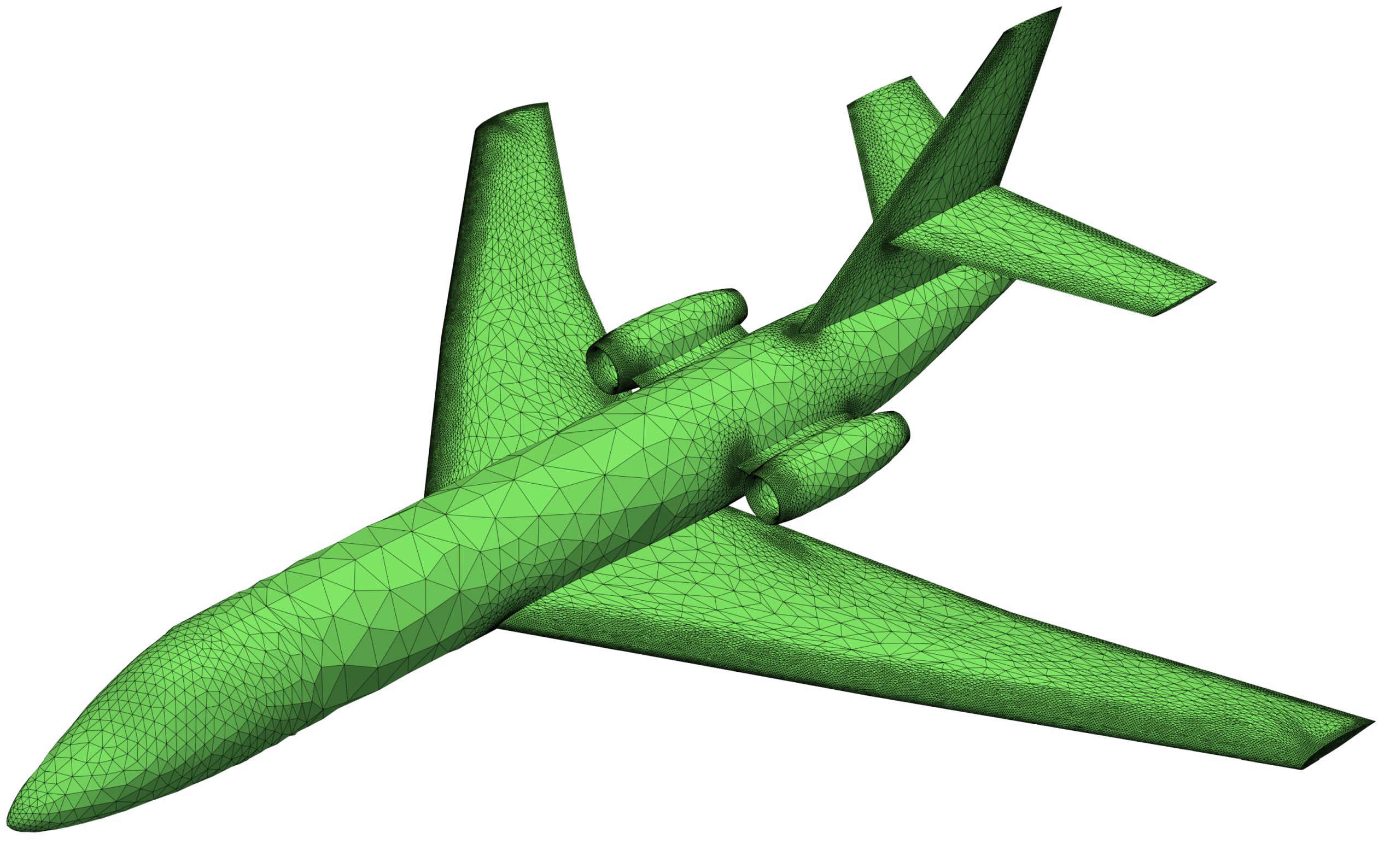}}
	\subfigure[Geometry 3]{\includegraphics[width=0.32\textwidth]{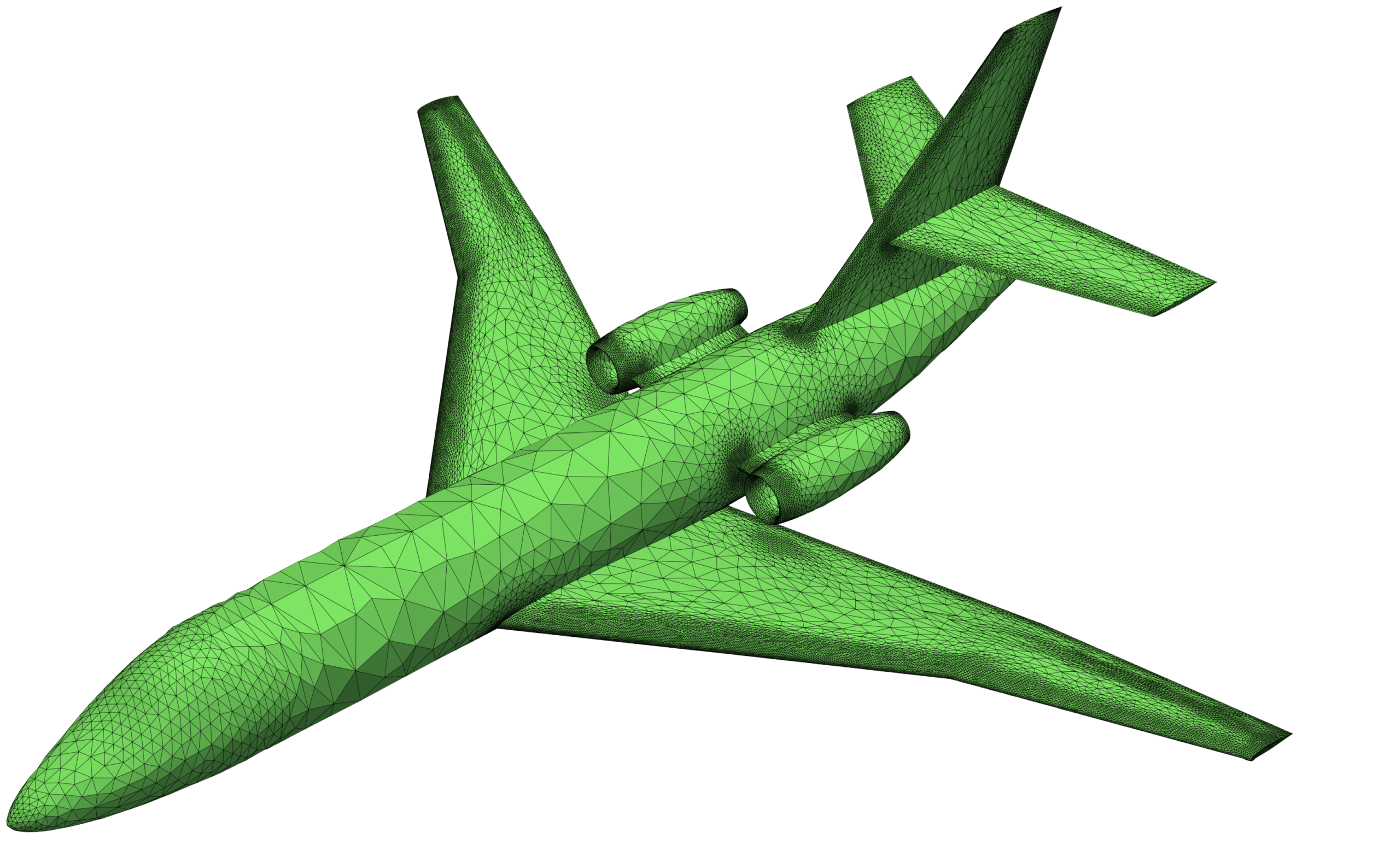}}
	\\
	\subfigure[Geometry 1]{\includegraphics[width=0.32\textwidth]{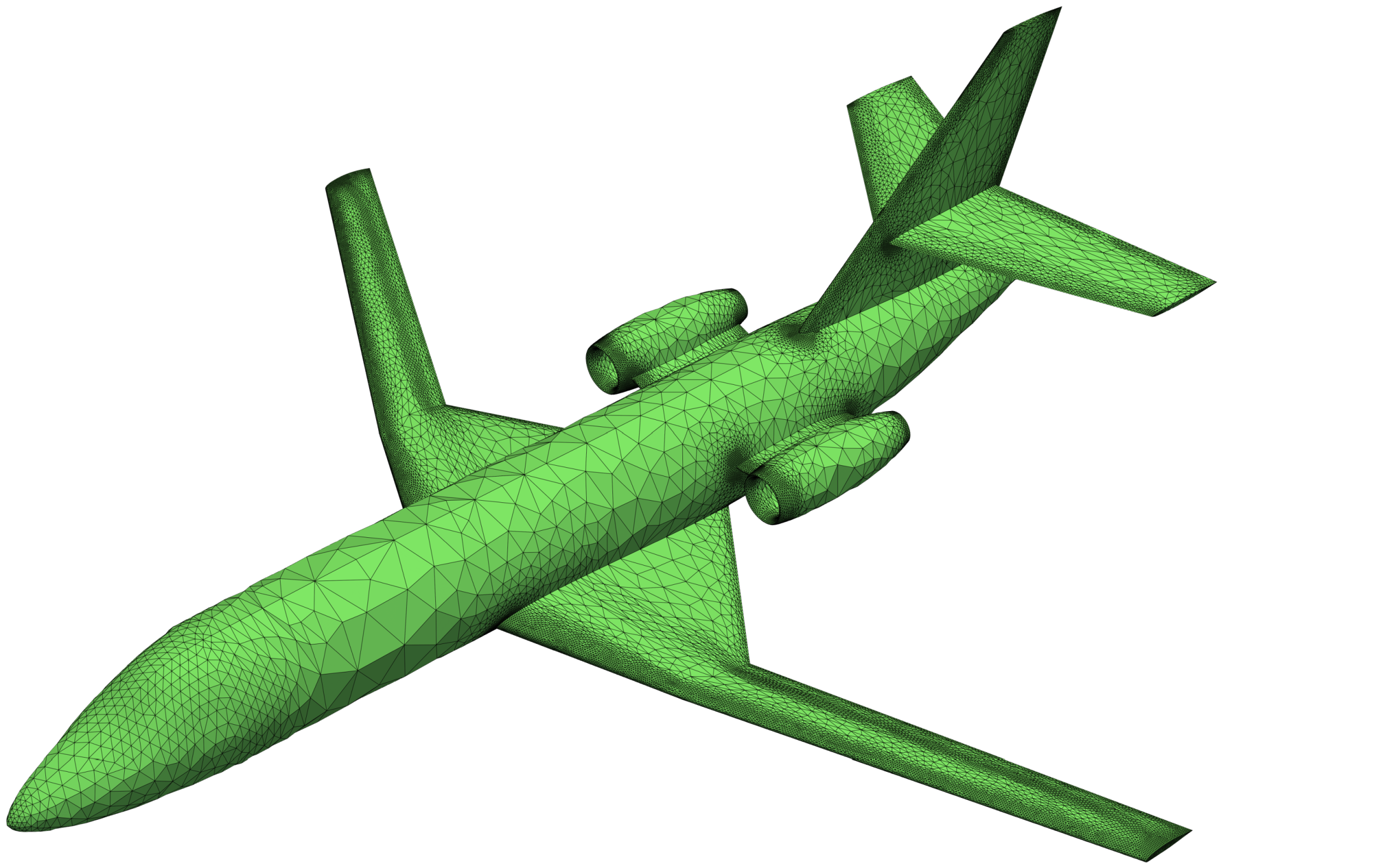}}
	\subfigure[Geometry 2]{\includegraphics[width=0.32\textwidth]{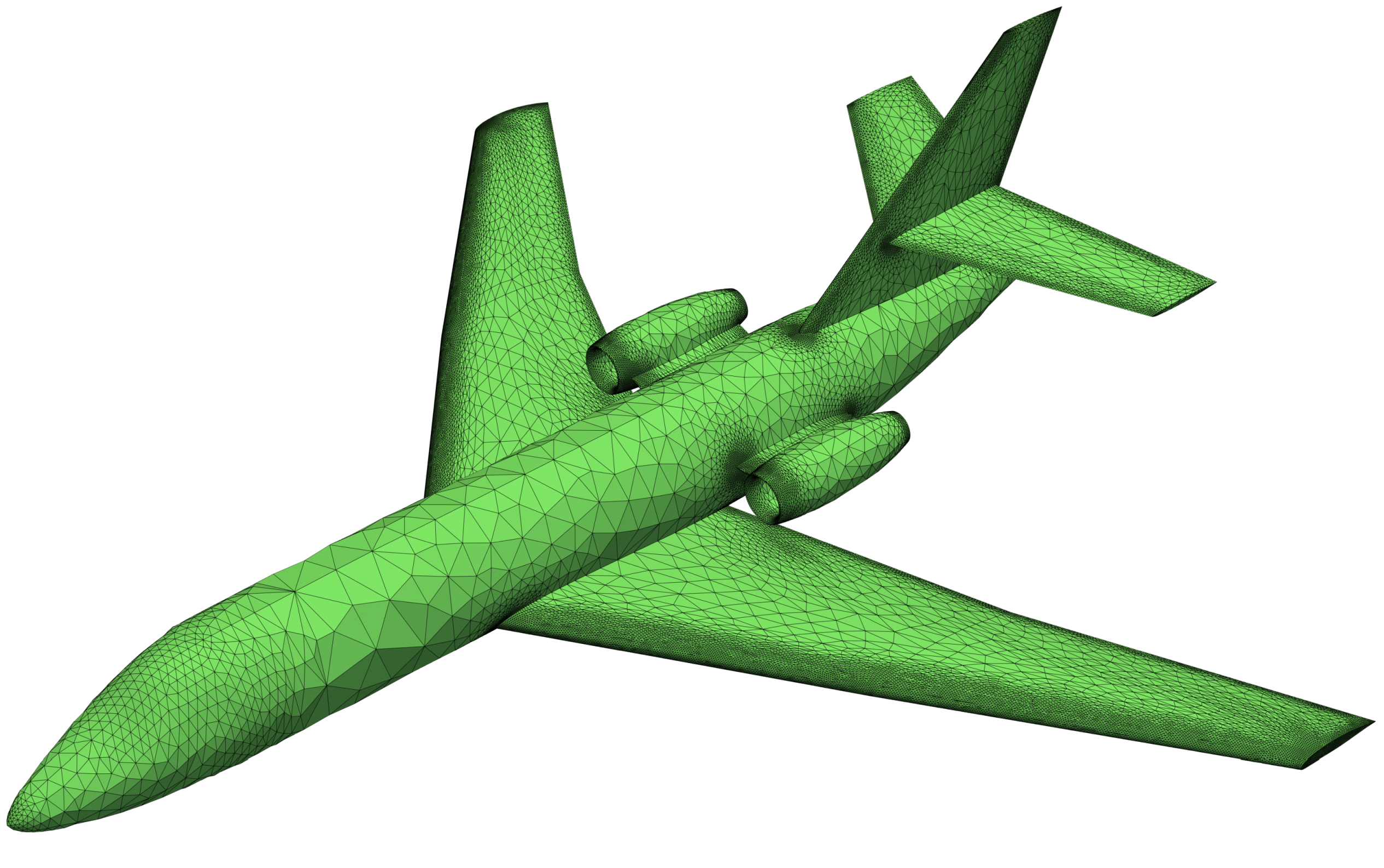}}
	\subfigure[Geometry 3]{\includegraphics[width=0.32\textwidth]{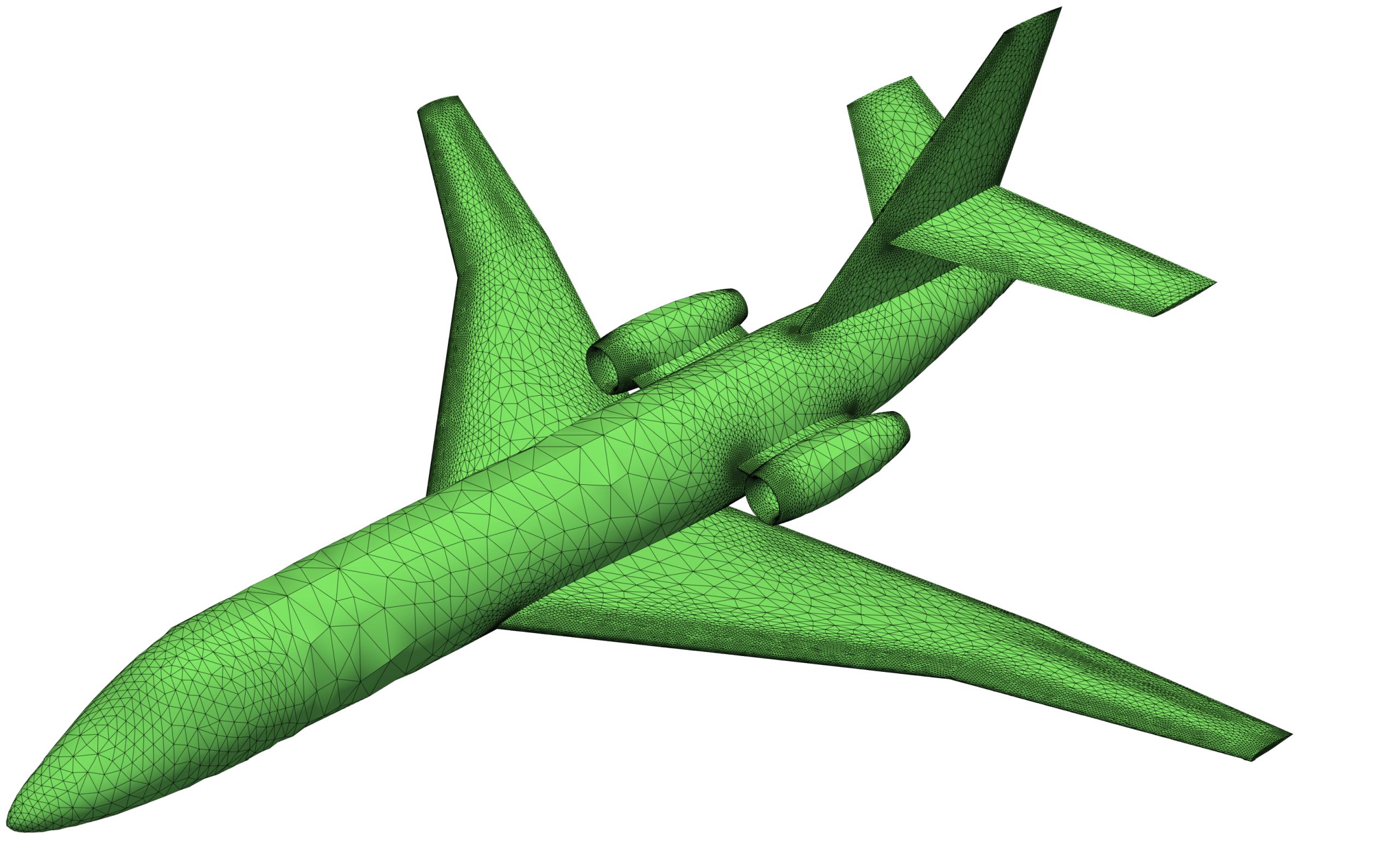}}
	\caption{Geometrically parametrised aircraft: Target (top) and predicted (bottom) meshes for three geometric configurations unseen by the ANN during training.}
	\label{fig:Falcon:flowTargetPredictionMesh} 
\end{figure}

To illustrate the ability of the proposed approach to predict anisotropic spacing for a more complex problem,  the trained ANNs are used to predict the metric tensor at each point of the domain, which is subsequently used to generate an anisotropic mesh for three cases unseen during training.  Figure~\ref{fig:Falcon:flowTargetPredictionMesh} compares the target and predicted meshes for the three cases. 
The results clearly show the ability of the trained ANN to accurately predict the regions where refinement is required as well as the anisotropic character of the target spacing.

\begin{figure}[!tb] 
	\centering
	\subfigure[Spacing in $\textbf{e}_1$]{\includegraphics[width=0.49\textwidth]{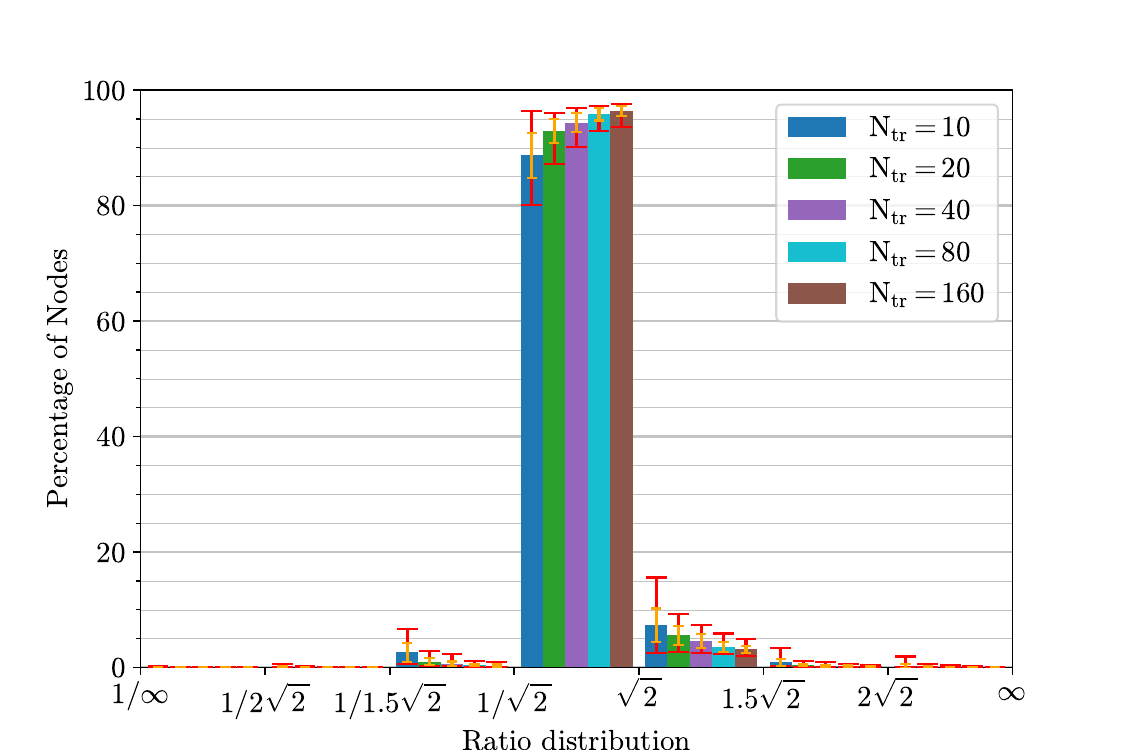}}
	\subfigure[Spacing in $\textbf{e}_2$]{\includegraphics[width=0.49\textwidth]{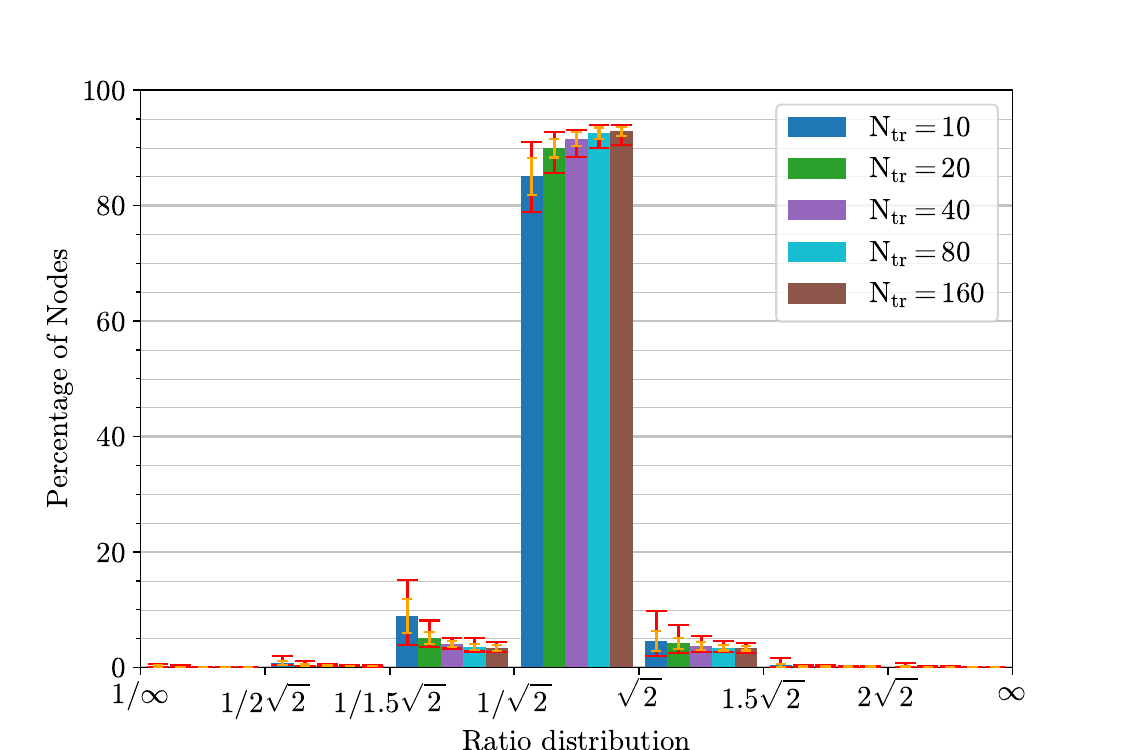}}
	\subfigure[Spacing in $\textbf{e}_3$]{\includegraphics[width=0.49\textwidth]{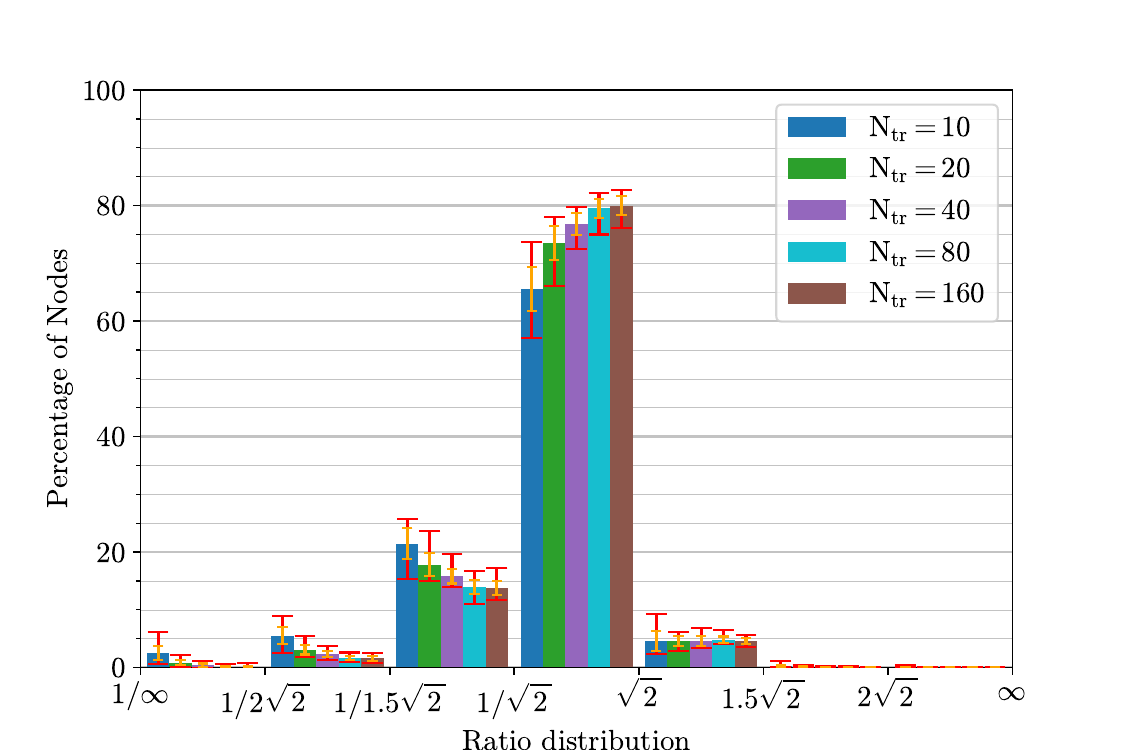}}
	\caption{Geometrically parametrised aircraft: Histogram of the ratio between the predicted and target spacings for an increasing number of training cases, $N_{tr}$.}
	\label{fig:Falcon:ModelSpacingComparison} 
\end{figure}

A more quantitative analysis is provided in Figure~\ref{fig:Falcon:ModelSpacingComparison}, showing the histogram of the anisotropic spacing accuracy for different training datasets.
The histograms show that with only 20 training cases it is possible to predict 90\% of the nodes with an acceptable spacing in the two first directions. It is remarkable that the accuracy in the first two directions is similar to the one obtained in the previous example, despite in this scenario 11 geometric parameters are considered and in the previous example only two flow conditions were used. The main difference between the two examples is that in the current example it is particularly challenging to predict the spacing in the third direction. However, it is worth noting that this direction corresponds to the maximum of the three spacings and therefore the least relevant. Furthermore, the histograms show that the trained ANN tends to refine more than required in this direction rather than under-refine, which is obviously preferable.

%%%%%%%%%%%%%%%%%%%%%%%%%%%%%%%%%%%%%%%%%%%%%%%%%%%%%%%%%%%%%%%%%%%%%%%%%%%%%%%%%%%%%%%%%%%%%%%%%%%%%%%
\section{Concluding remarks}\label{sc:conclusions}
%%%%%%%%%%%%%%%%%%%%%%%%%%%%%%%%%%%%%%%%%%%%%%%%%%%%%%%%%%%%%%%%%%%%%%%%%%%%%%%%%%%%%%%%%%%%%%%%%%%%%%%

A methodology to predict the near-optimal anisotropic spacing function for unseen simulations has been presented. The strategy aims to leverage the vast amount of  high-fidelity data that is normally available in industry to build an ANN capable of predicting anisotropic spacing.

The strategy involves computing the metric tensor at each mesh node and for each available simulation and transferring this information to a common coarse background mesh. The strategy to transfer the anisotropic spacing to the background meshes uses a conservative metric intersection approach aimed at minimising the possibility of producing meshes that miss key solution features.

Two ANN are trained independently to predict the directions of anisotropy and the corresponding spacings. The proposed ANN architecture exploits the mathematical properties of a metric tensor and predicts only two directions of anisotropy and three spacings. Different models were tested and compared to investigate if training different ANNs to predict the angles that define the first two anisotropic directions separately were beneficial. It was found that predicting the angles that define the two first directions of anisotropy using a single ANN provided the most accurate predictions.

Numerical examples involving three-dimensional inviscid compressible flow simulations are used to illustrate the potential of the proposed strategy and the accuracy of the predicted spacing functions. For an example with a fixed geometry and two parameters characterising the flow conditions, it was found that with only 10 training cases the ANN can predict an acceptable spacing in the first direction of anisotropy for almost 95\% of the nodes. It is worth remarking that the first direction of anisotropy is the most critical as it corresponds to the direction of minimum spacing. The predicted meshes were utilised to perform simulations and confirm that acceptable results can be obtained in terms quantities of interest such as lift and drag.

The second example, involving a full aircraft with 11 geometric parameters, was used to demonstrate the applicability in a more complex industrial setting. Despite the higher dimensionality and more complex flow features, the results showed that with 40 training cases it was possible to predict an acceptable spacing in the first direction of anisotropy for almost 95\% of the nodes.

%%%%%%%%%%%%%%%%%%%%%%%%%%%%%%%%%%%%%%%%%%%%%%%%%%%%%%%%%%%%%%%%%%%%%%%%%%%%%%%%%%%%%%%%%%%%%%%%%%%%%%%
\section*{Acknowledgements}
%%%%%%%%%%%%%%%%%%%%%%%%%%%%%%%%%%%%%%%%%%%%%%%%%%%%%%%%%%%%%%%%%%%%%%%%%%%%%%%%%%%%%%%%%%%%%%%%%%%%%%%

The authors are grateful for the financial support provided by the Engineering and Physical Sciences Research Council (EP/T517987/1).

%%%%%%%%%%%%%%%%%%%%%%%%%%%%%%%%%%%%%%%%%%%%%%%%%%%%%%%%%%%%%%%%%%%%%%%%%%%%%%%%%%%%%%%%%%%%%%%%%%%%%%%
\section*{Declaration of interest}
%%%%%%%%%%%%%%%%%%%%%%%%%%%%%%%%%%%%%%%%%%%%%%%%%%%%%%%%%%%%%%%%%%%%%%%%%%%%%%%%%%%%%%%%%%%%%%%%%%%%%%%

The authors have no competing interests to declare that are relevant to the content of this article.

\bibliographystyle{elsarticle-num} 
\bibliography{Paper_ref}

\end{document}